%% file: paper.tex
\documentclass{LMCS}

\def\dOi{9(3:18)2013}
\lmcsheading%
{\dOi}
{1--36}
{}
{}
{Jan.~15, 2013}
{Sep.~17, 2013}
{}

\usepackage{enumerate}
\usepackage{hyperref}

\usepackage[noadjust]{cite}
\usepackage{amsmath}
\usepackage{amssymb}

\usepackage{graphicx}

\usepackage{wrapfig}
\usepackage{comment}

\usepackage{sty/pi}
\usepackage{sty/pndraw}
\usepackage{sty/translation}
\usepackage{sty/mathematics}

\newcommand{\eg}             {e.g.\@\xspace}

\newcommand{\ie}             {i.e.\@\xspace}

\newcommand{\wrt}            {wrt.\@\xspace}
\newcommand{\viz}            {viz.\@\xspace}
\newcommand{\cf}             {cf.\@\xspace}
\newcommand{\wlogg}          {wlog.,\xspace}

\newcommand{\DEF}            {\stackrel{\mbox{{\tiny\rm df}}}{=}}
\newcommand{\nsymbol}        {\mathbb{N}}
\newcommand{\post}    [1]        {#1^\bullet}
\newcommand{\pre}     [1]        {{}^\bullet{#1}}
\newcommand{\STEP}     [1]       {\stackrel{#1}{\rightarrow}}

\newcommand{\sumProc}{\textstyle\sum_{i\in I}\pi_i.S_i}

\newcommand{\pc}{\mathop{|}}
\newcommand{\freeNames}[1]{\mathit{fn}\left(#1\right)}
\newcommand{\boundNames}[1]{\mathit{bn}\left(#1\right)}

\newcommand{\PUB} {\ensuremath{\mathcal{P}}\xspace} 
\newcommand{\RESTR} {\ensuremath{\mathcal{R}}\xspace} 
\newcommand{\NEW} {\ensuremath{\mathcal{N}}\xspace} 
\newcommand{\INP} {\ensuremath{\mathcal{I}}\xspace} 
\newcommand{\FRM} {\ensuremath{\mathcal{F}}\xspace} 

\newcommand{\pspace}{\textsc{PSPACE}}

\begin{document}

\title[A Polynomial Translation of $\pi$-Calculus FCPs to Safe
Petri Nets]{A Polynomial Translation of $\pi$-Calculus FCPs\\ to
Safe Petri Nets\rsuper*}

\author[V.~Khomenko]{Victor Khomenko\rsuper a}
\address{{\lsuper a}Newcastle University, UK}
\email{Victor.Khomenko@ncl.ac.uk}

\author[R.~Meyer]{Roland Meyer\rsuper b}
\address{{\lsuper{b,c}}University of Kaiserslautern, Germany}
\email{\{Meyer,Huechting\}@cs.uni-kl.de}

\author[R.~H\"uchting]{Reiner H\"uchting\rsuper c}
\address{\vskip-6 pt}

\thanks{This research was
supported by the \textsc{Epsrc} grants EP/G037809/1
(\textsc{Verdad}) and EP/K001698/1 (\textsc{Uncover}).}



\ACMCCS{[{\bf Theory of computation}]: Formal languages and automata
  theory; Logic---Verification by model checking; Computational
  complexity and cryptography---Problems, reductions and completeness;
  [{\bf Computing methodologies}]: Modeling and simulation---Model
  development and analysis---Model verification and validation; [{\bf
      Software and its engineering}]: Software organization and
  properties---Software system structures\,/\,Software functional
  properties}

\titlecomment{{\lsuper*}This is the full version of a paper presented at the
$23^\mathrm{rd}$ Int.\ Conf.\ on Concurrency Theory
(CONCUR'12).}

\begin{abstract}
\noindent We develop a polynomial translation from finite
control \picalc\ processes to safe low-level Petri nets. To our
knowledge, this is the first such translation. It is natural in
that there is a close correspondence between the control flows,
enjoys a bisimulation result, and is suitable for practical
model checking.
\end{abstract}

\maketitle

\section{Introduction}

Many contemporary systems -- be it software, hardware, or
network applications -- support functionality that
significantly increases their power, usability, and
flexibility:

\medskip

\noindent\textbf{Interaction}\quad Mobile systems permeate our
lives and are becoming ever more important. Indeed, the vision
of pervasive computing, where devices like mobile phones and
laptops are opportunistically engaged in interaction with a
user and with each other, is quickly becoming a reality.

\smallskip

\noindent\textbf{Reconfiguration}\quad Systems implement a
flexible interconnection structure, \eg cores in
Networks-on-Chip temporarily shut down to save power, resilient
systems continue to deliver (reduced) functionality even if
some of their modules develop faults, and ad-hoc networks
build-up and destroy connections between devices at runtime.

\smallskip

\noindent\textbf{Resource allocation}\quad Many systems maintain
multiple instances of the same resource (\eg network servers, or processor cores in a microchip)
that are dynamically allocated to tasks depending on the workload, power
mode, and priorities of the clients.

\medskip

The common feature of such systems is the possibility to form
\emph{dynamic} connections between individual modules.
It is implemented using reference passing: a module can become aware
of another module by receiving a reference (\eg in form of a network address)
to it.
This reference enables subsequent communication between the modules, and
can be understood as a dynamically created connection.
We will refer to this class of systems as \emph{Reference
Passing Systems (RPS)}.

As people are increasingly dependent on the correct
functionality of Reference Passing Systems, the costs incurred by design
errors can be extremely high.
However, even the conventional concurrent systems are
notoriously difficult to design correctly because of the
complexity of their synchronisation behaviour. Reference
passing adds another layer of complexity by making the
interconnect structure dynamic.
Hence, formal verification has to be employed to ensure the
correct behaviour of RPSs.

While the complexity of systems increases, the time-to-market is reducing.
To address this, system design has changed from a holistic to a compositional process:
the system is usually composed from pre-existing modules. This change in the
design process has to be mirrored in the verification process.
\emph{Verification has to focus on the inter-modular level rather than on
individual modules.}
Indeed, it is reasonable nowadays to assume that individual modules are
well-tested or formally verified by their vendors.
Moreover, the inter-module communication fabric (\eg a computer
network) is usually built of standard components and uses
standard protocols, and so can be assumed to be
correct.
This means bugs primarily arise in the interaction
between modules, and here verification is required to
ensure correctness of the \emph{system as a whole.}

When we verify the interaction between modules, we take advantage of this \emph{separation of verification concerns}.
We only have to model the modules' interfaces but can abstract away from their internal behaviour as well
as low-level communication concerns (\eg network behaviour), which we assume to be checked by
the vendor.
Traditionally, the interface behaviour is verified using re\-ly/gu\-a\-ran\-tee
techniques, potentially supported by theorem provers.
Due to undecidability reasons, however, theorem proving
generally requires substantial manual intervention to discharge
some of the proof obligations. In the interest of a short time
to market, our ambition is to lift fully automatic model
checking to the inter-modular level.

There are several formalisms suitable for modelling the interface behaviour of RPSs.
The main considerations in choosing such a formalism are its
expressiveness and the tractability of the associated
verification problems. Expressive formalisms like
\picalc~\cite{MilnerParrowWalker1992} and Ambient
Calculus~\cite{CG-98} are Turing powerful and so undecidable in
general. Fortunately, the ability to pass references
\textit{per se} does not lead to undecidability. One can impose
restrictions on the communication capabilities
\cite{AM2002,M2009}, control flow \cite{Dam1996,Pistore1999},
or interconnection shape \cite{M2009,M2008IFIP} to recover
decidability while retaining a reasonable modelling power.

\emph{Finite Control Processes (FCP)}~\cite{Dam1996} are a
fragment of \picalc\ where the system is modelled as a
parallel composition of a fixed number of sequential entities
(threads).
The control of each thread is represented by a
finite automaton, and the number of threads is bounded in
advance.
The threads communicate synchronously via channels
that they create, exchange and destroy at runtime.
These capabilities are often sufficient to faithfully model the interface behaviour of RPSs.
The appeal of FCPs is in combining this modelling power with
decidability of verification~\cite{Dam1996,MKS2009FI}.

In this paper, we contribute to FCP verification, following an
established approach. We translate the FCP under study into a
safe low-level Petri net (PN).
%
This translation bridges the gap between expressiveness and
verifiability:
While \picalc\ is suitable for modelling RPSs but difficult to
verify due to the complicated semantics, PNs are a low-level
formalism equipped with efficient analysis algorithms.
With the translation, all verification techniques and tools
that are available for PNs can be applied to analyse the
(translated) process.

%
\subsection{A polynomial translation}
\label{se-complexity_intro}
There is a large body of literature on \picalc\ to Petri net
translations (\cf Sect.~\ref{Section:RelatedWork} for a
detailed discussion).
Complexity-theoretic considerations, however, suggest that they
are all suboptimal for FCPs --- either in terms of size
\cite{MKS2009FI,BG2009,M2009,GM2009} or because of a too
powerful target formalism \cite{AM2002,KKN2006,DKK2008}.
The following shows that a polynomial translation of FCPs into
low-level safe PNs must exist.

Since the state of an FCP can always be described by a
string of length linear in the process size, an FCP can be
simulated by a Turing machine (TM) with a tape of linear
length.
Moreover, a TM with bounded tape can be modelled by a safe PN
of polynomial size (in the size of the control and the tape
length), see \eg~\cite{Esparza1998}.
Finally, a linear translation from safe PNs to FCPs is
described in Sect.~\ref{Section:PNtoFCPtranslation}.
That is, the three formalisms can simulate each other with only polynomial overhead.
This argument is in fact constructive, and shows
PSPACE-completeness of FCP model checking. Even more, it shows
that we can use safe PNs to verify FCPs. The problem with this
translation via TMs is that the safe PN resulting from an
FCP would be large and contrived, and thus of limited use
for practical verification.

These considerations motivated us to look for a direct
polynomial translation of FCPs to safe PNs, which is the main
contribution of this paper. We stress that our translation is
not just a theoretical result, but is also quite practical:
\begin{iteMize}{$\bullet$}
  \item it is natural in the sense that there is a strong
      correspondence between the control flow of the FCP
      and the resulting PN;
  \item the transition system of the FCP and that of its PN
      representation are bisimilar, which makes the latter
      suitable for checking temporal properties of the
      former;
  \item the resulting PN is compact (polynomial even in the
      worst case);
  \item we propose several optimisations that significantly reduce the PN's size in
      practice;
  \item we propose several extensions of the translation,
      in particular to polyadic communication, polyadic
      synchronisation, and match/mis\-match operators;
  \item experiments demonstrate that the translation
      is suitable for practical verification.
\end{iteMize}
Technically, our translation relies on three insights: (i) the
behaviour of an FCP $\new r.(S_1\parComp S_2)$ coincides with
the behaviour of $(S_1\{n/r\}\parComp S_2\{n/r\})$ where the
restricted name $r$ has been replaced by a fresh public name
$n$ (a set of fresh names that is linear in the size of the FCP
will be sufficient);
(ii) we have to recycle fresh names, and so implement reference
counters for them; and (iii) we hold substitutions explicit and
give them a compact representation using decomposition, \eg,
$\{a, b/x, y\}$ into $\{a/x\}$ and $\{b/y\}$.

\subsection{Related work}\label{Section:RelatedWork}

There are two main approaches to FCP verification. The first is
to directly generate the state space of the model, as is done
(on-the-fly) by the Mobility Workbench
(MWB)~\cite{VictorMoller1994}. This approach is relatively
straightforward but has a number of disadvantages. In
particular, its scalability is poor due to the complexity of
the \picalc semantics, which restricts the use of heuristics
for pruning the state space, and due to the need for expensive
operations (like equivalence checks \cite{KM2009}) every time a
new state is generated. Furthermore, some efficient model
checking techniques like symbolic representations of state
spaces are difficult to apply.

The alternative approach, and the one followed here, is to
translate a \picalc term into a simpler formalism, \eg a Petri
net, that is then analysed.
This method does not depend on a concrete verification technique for \picalc\ but can adapt any such
technique for PNs.
In particular, RPSs often are highly concurrent, and so
translating them into a true concurrency formalism like PNs
opens the door for partial-order reductions in verification.
This alleviates the problem of combinatorial state space
explosion, that is, a small specification often has a huge
number of reachable states that is beyond the capabilities of
existing computers.

Although several translations of \picalc\ to Petri nets have
been proposed, none of them provides a polynomial translation
of FCPs to safe PNs.
%
%
The verification kit
HAL~\cite{FerrariGnesiMontanariPistore2003} translates a model
into a \emph{History Dependent Automaton} --- a finite
automaton where states are labelled by sets of names that
represent restrictions~\cite{MontanariPistore19952,Pistore1999}.
For model checking, these automata are further translated to finite
automata \cite{FerrariGnesiMontanariPistore2003}.
Like in our approach, the idea is to replace restrictions with
fresh names, but the translation stores full substitutions,
which may yield an exponential blow up of the finite automaton.
Our translation avoids this blow up by compactly
representing substitutions by PN markings. This, however, needs
careful substitution manipulation and reference counting.

Amadio and Meyssonnier \cite{AM2002} replace unused restricted
names by generic free names. Their translation instantiates
substitutions, \eg $(\send{x_1}{y_1}.\send{x_2}{y_2})\{a, b, a,
b/x_1, y_1, x_2, y_2\}$ is represented by
$\send{a}{b}.\send{a}{b}$. This creates an exponential blow up:
since the substitutions change over time, $m$ public names and
$n$ variables may yield $m^n$ instantiated terms. Moreover,
since the number of processes to be modified by replacement is
not bounded in \cite{AM2002}, Amadio and Meyssonnier use PNs
with transfer. (Their translation handles a subset of \picalc\
incomparable with FCPs.) As this paper shows, transfer nets are
an unnecessarily powerful target formalism for FCPs
--- \eg reachability is undecidable in such nets~\cite{DufourdFinkelSchnoebelen1998}.

Busi and Gorrieri study non-interleaving and causal semantics
for the \picalc{} and provide decidability results for model
checking \cite{BG2009}. (That work has no bisimilarity proof,
which is provided in \cite{GM2009}.) The translations may be
exponential for FCPs, again due to the instantiation of
substitutions.

Devillers, Klaudel and Koutny \cite{DKK2008} achieve a
bisimilar translation of \picalc\ into high-level Petri nets,
thus using a Turing complete target formalism where automatic
analyses are necessarily incomplete. In \cite{KKN2006}, this
translation is used for unfolding-based model checking; to
avoid undecidability, the processes are restricted to be
recursion-free --- a class of limited practical applicability.

Peschanski, Klaudel and Devillers
\cite{PeschanskiKlaudelDevillers2011} translate $\pi$-graphs (a
graphical variant of \picalc) into high-level PNs. The
technique works on a fragment that is equivalent to FCPs.
However, the target formalism is unnecessarily powerful, and
the paper provides no experimental evaluation.

Our earlier translation \cite{M2009} identifies groups of
processes that share restricted names. In~\cite{MKS2009FI}, we
modify it to generate safe low-level PNs, and use an
unfolding-based model checker. The experiments indicate that
this technique is more scalable than the ones above, and it has
the advantage of generating low-level rather than high-level
PNs. However, the PN may still be exponentially large.

\section{Basic notions}\label{se-basics}

\subsection{Petri nets} A \emph{Petri net} (PN) is a tuple
$N\DEF(P,T,F,M_0)$ such that $P$ and $T$ are disjoint sets of
\emph{places} and \emph{transitions}, $F\subseteq (P\times
T)\cup(T\times P)$ is a \emph{flow relation}, and $M_0$ is the
\emph{initial marking} of $N$, where a \emph{marking}
$M:P\rightarrow\nsymbol\DEF\{0,1,2,\ldots\}$ is a
multiset of places in $N$. We draw PNs in the standard way: places are
represented as circles, transitions as boxes, the flow relation
by arcs, and a marking by tokens within circles. The
\emph{size} of $N$ is
\[
    \sizeApp{N}\DEF|P|+|T|+|F|+|M_0|.
\]

We denote by $\pre{z}\DEF\{y\mid(y,z)\in F\}$ and
$\post{z}\DEF\{y\mid(z,y)\in F\}$ the \emph{preset} and
\emph{postset} of $z\in P\cup T$, respectively. A transition
$t$ is \emph{enabled at marking $M$}, denoted by $M\STEP{t}$,
if $M(p)>0$ for every $p\in\pre{t}$. Such a transition can
\emph{fire}, leading to the marking $M'$ with
\[
    M'(p)\DEF M(p)-F(p,t)+F(t,p)\quad\text{for every }p\in P.
\]
We denote the firing
relation by $M\STEP{t}M'$ or by $M\rightarrow M'$ if the
identity of the transition is irrelevant. The set of
\emph{reachable markings of $N$} is denoted by $\reachof{N}$.
The \emph{transition system} of $N$ is
\begin{align*}
\tsof{N}\DEF
(\reachof{N},\rightarrow,M_0).
\end{align*}

A PN $N$ is \emph{$k$-bounded} if $M(p)\leq k$ for every
$M\in\reachof{N}$ and every place $p\in P$, and \emph{safe} if
it is $1$-bounded. This paper focuses on safe PNs. A set of
places in a PN is \emph{mutually exclusive} if at any reachable
marking at most one of them contains tokens. A place $p$ is a
\emph{complement} of a set $Q$ of mutually exclusive places if
at any reachable marking $p$ contains a token iff none of the
places in $Q$ contains a token. If $Q=\{q\}$, the places $p$
and $q$ are \emph{complements} of each other.

\subsection{Finite control processes} In \picalc
\cite{Milner1999,SangiorgiWalker2001}, threads communicate via
synchronous message exchange. The key idea in the model is that
messages and the channels they are sent on have the same type:
they are just \emph{names} from some set $\names\DEF\{a,b,x, y,
i, f, r, \ldots\}$. This means a name that has been received as
a message in one communication may serve as a channel in a
later interaction. To communicate, processes consume
\emph{prefixes}
\begin{align*}
\pi::=\send{a}{b}\bnf \rec{a}{x}\bnf \tau.
\end{align*}
The \emph{output prefix} $\send{a}{b}$ sends name $b$ along
channel $a$. The \emph{input prefix} $\rec{a}{x}$ receives a
name that replaces $x$ on channel $a$. Prefix $\tau$ stands for
a \emph{silent action.}

\emph{Threads}, also called \emph{sequential processes}, are constructed as
follows.
A \emph{choice process} $\sumProc$ over a finite set of indices
$I$ executes a prefix $\pi_i$ and then behaves like $S_i$. The
special case of choices over an empty index set $I=\emptyset$
is denoted by $\stopProc$. This process terminates a thread.
A \emph{restriction} $\new r.S$ generates a name $r$ that is
different from all other names in the system. To implement
parameterised recursion, we use \emph{calls to process
identifiers} $\call{K}{\tilde a}$. We defer their explanation
for the moment. To sum up, threads take the form
\begin{align*}
S::={\call{K}{\tilde a}}
	\bnf \sumProc
	\bnf{\new r.S}.
\end{align*}
We use $\sProc$ to refer to the set of all threads. A
\emph{finite control process (FCP) $F$} is a \emph{parallel
composition} of a fixed number of threads:
\begin{align*}
 F::= \new \tilde r.(S_{\mathit{init, 1}}\parComp \ldots \parComp
S_{\mathit{init, n}}).
\end{align*}
Here, $\new \tilde r$ with $\tilde r=r_1\ldots r_k$ denotes a
(perhaps empty) sequence of restrictions $\new r_1\ldots
\new r_k$.
We use $P$ to refer to an arbitrary process, be it a thread, an
FCP, or a term obtained by structural congruence defined below.
We denote iterated parallel compositions by $\prod_{i\in I}P_i$.

Our presentation of parameterised recursion using calls
$\call{K}{\tilde a}$ follows \cite{SangiorgiWalker2001}.
Process identifiers $K$ are taken from some set
$\PIDS\DEF\{H,K,L,\ldots\}$ and have a \emph{defining equation}
$K(\tilde f):=S$. Thread $S$ can be understood as the
implementation of $K$. The process has a list of distinct
\emph{formal parameters} $\tilde f=f_1,\ldots, f_k$ that are
replaced by \emph{factual parameters} $\tilde a=a_1,\ldots,
a_k$ when a call $\call{K}{\tilde a}$ is executed. Note that
$\tilde a$ and $\tilde f$ have the same length.
When talking about an \emph{FCP specification} $F$, we mean
process $F$ with all its defining equations.

To implement the replacement of formal parameters $\tilde f$ by
$\tilde a$ in calls to process identifiers, we use
\emph{substitutions}. A substitution is a function
$\sigma:\names\rightarrow \names$ that maps names to names. If
we make domain and codomain explicit, $\sigma:A\rightarrow B$
with $A, B\subseteq \names$, we require $\sigma(a)\in B$ for
all $a\in A$ and $\sigma(x)= x$ for all $x\in\names\setminus
A$. We use $\{\tilde a/\tilde f\}$ to denote the substitution
$\sigma:\tilde f\rightarrow \tilde a$ with $\sigma(f_i)\DEF
a_i$ for $i\in\{1,\ldots, k\}$. The \emph{application of
substitution $\sigma$ to $S$} is denoted by $S\sigma$ and
defined in the standard way~\cite{SangiorgiWalker2001}.

Input prefix $\rec{a}{i}$ and restriction $\new r$ \emph{bind}
the names $i$ and $r$, respectively. The \emph{set of bound
names} in a process $P$ is $\boundNames{P}$. A name
which is not bound is \emph{free}, and the \emph{set of free
names} in $P$ is $\freeNames{P}$. We permit \alphaConv{} of
bound names. Therefore, \wlogg we make the following
assumptions common in \picalc theory and collectively referred
to as \emph{no clash} \noclash\ henceforth. For every FCP
specification $F$, we require that: (i) a name is bound at most
once, a name $f$ is used at most once in a formal parameter
list, bound and free names are disjoint, bound names and formal
parameters are disjoint, formal parameters and free names in
$F$ are disjoint; and (ii) if $\sigma=\{\tilde a/\tilde x\}$ is
applied to $S$ then $\boundNames{S}\cap(\tilde a\cup\tilde
x)=\emptyset$.

Assuming \noclash, the names in an FCP specification $F$
and the corresponding defining equations can be partitioned
into the following sets: set $\RESTR$ of names bound by
restriction operators, set $\INP$ of names bound by input
prefixes, set $\FRM$ of names used as formal parameters in
defining equations, and set $\PUB$ of all the other names ---
they are called \emph{public}.

We are interested in the relation between the size of an FCP
specification and the size of its PN representation. The
\emph{size} of an FCP specification is defined as the size of
its initial term plus the sizes of the defining equations:
\vspace{0.2cm}
\begin{center}
$\sizeApp{\stopProc}\DEF 1\qquad\sizeApp{\sumProc}\DEF
3\power{I}{-}1{+}\textstyle\sum_{i\in
I}\sizeApp{S_i}\qquad\sizeApp{\new r.P}\DEF
1{+}\sizeApp{P}\qquad \sizeApp{\call{K}{\tilde a}}\DEF
1{+}\power{\tilde a}$\\
$\sizeApp{S_{\init, 1}\parComp\ldots\parComp S_{\init, n}}\DEF
n{-}1{+}\sum_{1\leq i\leq n}\sizeApp{S_{\init,
i}}\qquad\sizeApp{K(\tilde f):=S}\DEF 1{+}\power{\tilde
f}{+}\sizeApp{S}$.
\end{center}
\vspace{0.2cm} Intuitively, we count the operations, process
identifiers, and names in each element of the specification,
\eg the factor $3$ in $3\power{I}$ refers to a send or receive
prefix of size two, followed by a plus.

\begin{figure}[t]
\centering
    \fbox{
    $
    \begin{array}{c}
    \new r.P\ \structCong\ \new r'.P\{r'/r\}\mbox{ if $r'\notin\freeNames{P}$}
    \quad\quad
    \rec{a}{x}.S\ \structCong\ \rec{a}{x'}.S\{x'/x\}\mbox{ if $x'\notin\freeNames{S}$}\medskip\\[0.2cm]
    P_1\parComp P_2\ \structCong\ P_2\parComp P_1
    \quad\quad
    (P_1\parComp P_2)\parComp P_3\ \structCong\ P_1\parComp (P_2\parComp P_3)
    \quad\quad
    P\parComp \stopProc\ \structCong\ P\medskip\\[0.2cm]
    {\new r.}\stopProc\ \structCong\ \stopProc
    \quad\quad
    \new r_1.\new r_2.P\ \structCong\ \new r_2.\new r_1.P
    \quad\quad
    \new r.(P_1 \parComp P_2)\ \structCong\ P_1 \parComp \new r.P_2\mbox{ if $r\notin \freeNames{P_1}$}
    \end{array}
    $
    }
\caption{Axioms defining structural congruence $\structCong$.}
\label{fi-congruence-axioms}
\end{figure}

\begin{figure}[t]
\centering
    \fbox{
    $
    \begin{array}{c}
        \mbox{(Tau)~~}\tau.S+\ldots\rightarrow S
        \qquad
        \mbox{(Const)~~}\call{K}{\tilde a}\rightarrow S\{\tilde a/\tilde x\}\mbox{ if $K(\tilde x):=S$}\\[0.2cm]
        \mbox{(React)~~}\rec{x}{y}.S_1+\ldots\parComp\send{x}{z}.S_2+\ldots\rightarrow S_1\{z/y\}\parComp S_2\\[0.2cm]
        \mbox{(Par)~~}\dfrac{P_1\rightarrow P_1'}{P_1\parComp P_2\rightarrow P_1'\parComp P_2}
        \qquad
        \mbox{(Res)~~}\dfrac{P\rightarrow P'}{\new a.P\rightarrow \new a.P'}
        \qquad
        \mbox{(Struct)~~}\dfrac{P_1\structCong P_1'\rightarrow P_2'\structCong P_2}{P_1\rightarrow P_2}
    \end{array}
    $
    }
\caption{Axioms and rules defining reaction relation $\rightarrow$.}
\label{fi-reaction-relation}
\end{figure}

To define the behaviour of a process, we rely on the
\emph{structural congruence} relation $\structCong$ on process
terms. It is the smallest congruence satisfying the axioms in
Fig.~\ref{fi-congruence-axioms}, \ie \alphaConv{} of bound
names is allowed, ${\pc}$ is commutative and associative with
$\stopProc$ as the neutral element, and for choices
associativity and commutativity are enforced by the notation
$\sumProc$, $\new$ can be eliminated when applied to
$\stopProc$, is commutative, and its scope can be extended to
include a concurrent process not containing free occurrences of
the bound name.

The behaviour of \picalc processes is determined by
\emph{reaction relation} $\rightarrow$ between
terms~\cite{Milner1999,SangiorgiWalker2001}, see
Fig~\ref{fi-reaction-relation}. It has the axioms for consuming
silent prefixes, identifier calls and communications, and is
defined to be closed under $\pc$, $\new$, and $\structCong$.
By $\reachof{F}$ we denote the set of \emph{all processes
reachable from $F$}. The \emph{transition system} of $F$
factorises the reachable processes along structural congruence,
\begin{align*}
\tsof{F}\DEF(\factorize{\reachof{F}}{\structCong},
\rightarrowts, \class{F}),
\end{align*}
where $\class{F}$ denotes the congruence class of $F$ \wrt
$\structCong$ and $\class{F_1}\rightarrowts\class{F_2}$ if
$F_1\rightarrow F_2$. That is, structurally congruent processes
are collapsed into a single state, and the transition relation
is amended appropriately.

\subsubsection*{Normal form assumptions}

To ease the definition of the translation and the corresponding
correctness proofs, we make assumptions about the shape of the
FCP specification. These assumptions are not restrictive, as
any FCP can be translated into the required form. First, we
require that the sets of identifiers called by different
threads (both directly from $F$ and indirectly from defining
equations) are disjoint. This ensures that the threads have
disjoint descriptions of their control flows and corresponds to
the notion of a \emph{safe} FCP in~\cite{MKS2009FI}. The
assumption can be achieved by replicating some defining
equations. The resulting specification $F'$ is bisimilar with
$F$ and has size $O(n\sizeApp{F})=O(\sizeApp{F}^2)$, where $n$
is the number of threads. We illustrate the construction on the
FCP specification $F = \call{K}{a,b} \pc \call{L}{a,c}$ (left)
together with its replicated version $F' = \call{K^1}{a,b} \pc
\call{L^2}{a,c}$ (right):
\begin{align*}
    K(f_1,f_2)&:=\tau.\call{L}{f_1,f_2} &\hspace{1.2cm} K^1(f_1^1,f_2^1)&:=\tau.\call{L^1}{f_1^1,f_2^1}\\
    L(f_3,f_4)&:=\tau.\call{K}{f_3,f_4} & L^1(f_3^1,f_4^1)&:=\tau.\call{K^1}{f_3^1,f_4^1}\\
    && K^2(f_1^2,f_2^2)&:=\tau.\call{L^2}{f_1^2,f_2^2}\\
    && L^2(f_3^2,f_4^2)&:=\tau.\call{K^2}{f_3^2,f_4^2}.
\end{align*}
We can also ensure that defining equations do not call
themselves, \ie that the body of $K(\tilde f):=S$ does not
contain any calls of the form $\call{K}{\tilde a}$. Indeed, we
can replace any such call with $\call{K'}{\tilde a}$, using a
new defining equation $K'(\tilde f'):=\call{K}{\tilde f'}$.
This increases the size of the FCP only linearly, and ensures
we do not have to remap parts of $\tilde f$ to $\tilde f$ when
passing the parameters of a call, which simplifies the
translation.

\section{Principles of the translation}\label{se-translation}
This section informally explains the polynomial translation of
FCPs to safe PNs.
The main idea is to replace restricted names by fresh public ones.
Indeed, $F = \new \tilde r.(S_{1}\parComp \ldots\parComp
S_{n})$ behaves like $S_{1}\{\tilde n/\tilde
r\}\parComp\ldots\parComp S_{n}\{\tilde n/\tilde r\}$, provided
the names $\tilde n$ are fresh.
These new names are picked from a set \NEW, and since for FCP
specifications there is a bound on the number of restricted
names in all processes reachable from $F$, a finite \NEW
suffices.
%
%
But how to support name creation and deletion with a constant
number of free names?
The trick is to reuse the names: $n\in\NEW$ may first represent a restricted name $r_1$ and
later a different restricted name $r_2$.
To implement this recycling of names, we keep track of whether
or not $n\in\NEW$ is currently used in the process. This can
be understood as reference counting.

The translation takes the set of names $\NEW$ as a parameter.
Already a fairly large set $\NEW_{\RESTR\INP\FRM}$ of
cardinality $\power{\RESTR}+\power{\INP}+\power{\FRM}$ is
sufficient to prove polynomiality of the translation. The
rationale is that there should be enough values to map each
bound name to a unique value. Indeed, one can provide a
\emph{domain function}
$\dom:\PUB\cup\RESTR\cup\INP\cup\FRM\to2^{\PUB\cup\NEW}$ that
gives, for each name $x$ of $F$, an overapproximation of the
set of possible values of $x$. The rough overapproximation
above employs
\begin{align*}
    \dom_{\RESTR\INP\FRM}(x)
    \DEF
    \left\{
        \begin{array}{l@{\qquad}l}
          \{x\} & \mbox{if $x\in\PUB$} \\
          \NEW_{\RESTR\INP\FRM} & \mbox{if $x\in\RESTR$} \\
          \PUB\cup\NEW_{\RESTR\INP\FRM} & \mbox{if $x\in\INP\cup\FRM$.} \\
        \end{array}
    \right.
\end{align*}
We explain how to compute better domains by static analysis in
Sect.~\ref{se-opt}.

The translation is then defined to be the composition
\begin{align*}
N(F) \DEF \nsubst \impl H(N(S_{1})\parallel\ldots \parallel N(S_{n})).
\end{align*}
The first net $\nsubst$ compactly represents substitutions
$\sigma:\RESTR\cup\INP\cup\FRM\to\PUB\cup\NEW$ and implements
reference counting. It only consists of places and is detailed
in Sect.~\ref{se-subst}. The second net
$H(N(S_{1})\parallel\ldots
\parallel N(S_{n}))$ represents the control flow of the FCP.
Each net $N(S_i)$ is a finite automaton that reflects the
control flow of thread $S_i$. Importantly, the transitions of
$N(S_i)$ are annotated with synchronisation actions and sets of
commands that explicitly handle the introduction and removal of
name bindings. Parallel composition $\parallel$ synchronises
the subnets of all threads. The operator places the nets side
by side and merges pairs of transitions with complementary
synchronisation actions $\mathit{send}(a, b)$ and
$\mathit{rec}(a, b)$. Hiding $H$ then removes the original
transitions. After hiding, transitions in the control flow net
only carry name binding commands. The implementation operator
$\impl$ implements them by adding arcs between the control flow
net and $\nsubst$. We elaborate on the construction of the
control flow in Sect.~\ref{se-control}, and
Sect.~\ref{se-example} illustrates the translation on an
example.

\subsection{Petri net representation of substitutions}\label{se-subst}

A substitution $\sigma:\RESTR\cup\INP\cup\FRM\to\PUB\cup\NEW$
maps the bound names and formal parameters occurring in the FCP
to their values. The compact PN representation of substitutions
is a key element of the proposed translation. It should
efficiently support the following operations:\medskip

\noindent\textbf{Initialisation of a restricted name}\quad It
should be possible to find a value $\mathit{val}\in\NEW$ to
which no bound name or formal parameter is currently mapped,
and map a given restricted name $r$ to $\mathit{val}$.\medskip

\noindent\textbf{Remapping}\quad When name $v$ is communicated
to a thread, some input name $i$ has to be mapped to
$\sigma(v)$. Similarly, a formal parameter $f$ has to be mapped
to $\sigma(v)$ during a process call that uses $v$ as a factual
parameter. Since $v$ can occur several times in the list of
factual parameters, one should be able to map several formal
parameters to $\sigma(v)$ in one step, \ie by one PN
transition.\medskip

\noindent\textbf{Unmapping}\quad When a bound name or formal
parameter $v$ is forgotten, its mapping has to be removed.
During a process call, it often happens that $\sigma(v)$ is
assigned to one or more formal parameters and simultaneously
$v$ is forgotten. Therefore, it is convenient to be able to
remap and unmap $v$ in one step.\medskip

Ideally, the three kinds of operations described above should
not interfere when applied to names in distinct threads, so
that they can be performed concurrently; note that due to
\noclash{} the bound names and formal parameters are always
different. This prevents the introduction of arbitration, and
so has a beneficial effect on the performance of some model
checking methods (\eg those using partial order
techniques).\medskip

\input{pics/substitution_net}

In what follows, we describe a representation of substitutions
that satisfies all the formulated requirements. This safe PN,
which is depicted in Fig.~\ref{fi-subst}, only consists of
places. A place $[\mathit{var}=\mathit{val}]$, when marked,
represents the fact that
$\mathit{var}\in\RESTR\cup\INP\cup\FRM$ is mapped to
$\mathit{val}\in\PUB\cup\NEW$. Our translation will maintain
the following invariants. (i) For each
$\mathit{var}\in\INP\cup\FRM$ and $\mathit{val}\in\NEW$, place
$[var\neq\mathit{val}]$ is complementary to
$[\mathit{var}=\mathit{val}]$. (ii) For each
$\mathit{val}\in\NEW$, the places
$[r_1=\mathit{val}],[r_2=\mathit{val}],\ldots$ are mutually
exclusive, so that no two restricted names can be mapped to the
same value. Moreover, $[r_*\neq\mathit{val}]$ is complementary
to all these places. (iii) For each
$\mathit{var}\in\RESTR\cup\INP\cup\FRM$, the places
$[\mathit{var}=\mathit{val}]$ are mutually exclusive (\ie a
name can be mapped to at most one value), where $\mathit{val}$
runs through $\dom(var)$.

The choice of the cardinality of $\NEW$ is of crucial
importance. As explained above, it should be sufficiently large
to guarantee that there will always be a name that can be used
to initialise a restriction. But taking an unnecessary big
value for this parameter increases the size of the PN as well
as the number of reachable states.

The operations on the substitution are implemented as
follows:\medskip

\noindent\textbf{Initialisation of a restricted name}\quad To
find a value $\mathit{val}\in\NEW$ that is not referenced, and
to map a given restricted name $r$ to $\mathit{val}$, the
transition has to:
\begin{iteMize}{$\bullet$}
  \item test by read arcs that the places
      $[i_1\neq\mathit{val}],[i_2\neq\mathit{val}],\ldots$
      and
      $[f_1\neq\mathit{val}],[f_2\neq\mathit{val}],\ldots$
      have tokens (\ie no input name or formal parameter is
      currently mapped to $\mathit{val}$);
  \item consume the token from $[r_*\neq\mathit{val}]$
      (checking thus that no restricted name is currently
      mapped to $\mathit{val}$);
  \item produce a token at $[r=\mathit{val}]$ (mapping thus
      $r$ to $\mathit{val}$).\medskip
\end{iteMize}

\noindent\textbf{Remapping}\quad When a communication binds the
value of a name~$v$ to an input name~$i$, the corresponding
transition consumes the token from $[i\neq\sigma(v)]$ (provided
$\sigma(v)\in\NEW$) and produces a token in $[i=\sigma(v)]$. In
the case of identifier calls, $v$ may occur several times in
the list of factual parameters, and so several formal
parameters $f_{k_1},\ldots,f_{k_l}$ have to be bound to
$\sigma(v)$. This can be handled by a single transition
consuming the tokens from $[f_{k_j}\neq\sigma(v)]$ and
producing tokens in $[f_{k_j}=\sigma(v)]$ for $j\in\{1,\ldots,
l\}$. The same transition can unmap $v$ if necessary.\medskip

\noindent\textbf{Unmapping}\quad When a bound name or formal
parameter $\mathit{var}$ that is mapped to $\mathit{val}$ is
forgotten, the mapping should be removed. This is modelled by a
transition consuming the token from
$[\mathit{var}=\mathit{val}]$ and, if $\mathit{val}\in\NEW$,
producing a token in $[\mathit{var}\neq\mathit{val}]$ (if
$\mathit{var}\in\INP\cup\FRM$) or
$[\mathit{var}_*\neq\mathit{val}]$ (if
$\mathit{var}\in\RESTR$).


\subsection{Petri net representation of the control flow}\label{se-control}
We elaborate on the translation of a thread $S_{i}$ into the net $N(S_{i})$ that reflects the control flow.
Each subterm $st$ of $S_{i}$ corresponds to a subnet of $N(S_{i})$ with a unique entry place $p_{st}$.
This place is initially marked if $st$ corresponds to the thread's initial expression.
The communication prefixes are modelled by transitions.
These transitions are labelled with synchronisation actions as well as commands to explicitly introduce and remove name bindings in the substitution net.
At this point, however, the transitions are just stubs.
The synchronisation between threads is introduced by parallel composition $\parallel$, the synchronisation between threads and substitution net is performed by the implementation operator $\impl$.

The subnets corresponding to terms $st$ are defined as follows
(note that each thread is a sequential process, so $\pc$ does
not occur in the term):\medskip

\noindent\textbf{Initialisation of restricted names}\quad The
subnet corresponding to $\new r.S$ maps $r$ to some currently
unused $n\in\NEW$, \cf Fig.~\ref{fi-restr}. More precisely, for
each $n\in\NEW$ we create a transition $t^r_n$ consuming a
token from the entry place of $\new r.S$, producing a token in
the entry place of the subnet implementing $S$, and performing
the initialisation of the restricted name $r$ with the value
$n$ as explained in Sect.~\ref{se-subst}. Note that the
transitions $t^r_n$ arbitrate between the names in $\NEW$,
allowing any of the currently unused names to be selected for
the initialisation of $r$. If such an arbitration is
undesirable,\footnote{E.g.\ due to its negative impact on some
model checking techniques. Note however that symmetry reduction
mitigates this negative effect, as all the states that are
reached by the arbitration are equivalent.} separate pools of
values can be used for each thread, as described in
Sect.~\ref{se-opt}.\medskip

\noindent\textbf{Handling $\stopProc$}\quad The subnet for
$\stopProc$ is comprised of the entry place only, which means
an execution of $\stopProc$ terminates a thread. An alternative
for implementing termination is to also unmap all the bound
names and formal parameters in whose scope this $\stopProc$
resides. We note that this unmapping is not necessary, as the
used resources (in particular, the values from \NEW to which
these variables are mapped) will not be needed.\medskip

\noindent\textbf{Handling calls}\quad Consider $\call{K}{\tilde
a}$ with $K(\tilde f):=S$. The entry place of $\call{K}{\tilde
a}$ is followed by a subnet that maps $\tilde f$ to the values
of $\tilde a$. All the other bound names and formal parameters
in whose scope this call resides become unmapped. Finally, the
control is transferred to the entry place of the translation
of~$S$.

To make this idea precise, let $B$ denote the set of bound names and formal
parameters that are forgotten in the call.
Let $A$ be the set of names occurring in $\tilde a$ (perhaps multiple times).
The required change in the substitution can be modelled by the assignments\vspace{0.2cm}
\begin{center}
    $X_i\leftarrow a$ for each $a\in A$
    \qquad and \qquad
    $\emptyset\leftarrow a$ for each $a\in B\setminus A$.\vspace{0.2cm}
\end{center}
Here, $X_i$ is the set of formal parameters to which the value
of the factual parameter $a$ is assigned. So the $X_i$ are
disjoint non-empty sets whose union is $\tilde f$. An
assignment $X\leftarrow a$ (where $X$ can be empty)
simultaneously maps all the variables in $X$ to $\sigma(a)$
and, if $a\in B$, unmaps $a$. Since no two assignments
reference the same name, they cannot interfere and thus can be
executed in any order or concurrently.

The subnet implementing an assignment $X\leftarrow a$ has one
entry and one exit place and is constructed as follows. For
each $\mathit{val}\in\dom(a)$ we create a transition
$t_\mathit{val}$ which:
\begin{iteMize}{$\bullet$}
  \item consumes a token from the entry place and produces
      a token on the exit place;
  \item for each $f\in X$, consumes a token from
      $[f\neq\mathit{val}]$ (provided this place exists,
      \ie $\mathit{val}\in\NEW$) and produces a token on
      $[f=\mathit{val}]$;
  \item if $a\in B$, consumes a token from $[a=\mathit{val}]$, and, in case
      $\mathit{val}\in\NEW$, produces a token on $[a\neq\mathit{val}]$ (or on
      $[r_*\neq\mathit{val}]$ if $a$ is a restricted
      name).
\end{iteMize}
Such subnets can be combined in either sequential or parallel
manner (in the latter case additional fork and join transitions
are needed).\medskip

\noindent\textbf{Handling sums}\quad We assume that sums are
\emph{guarded,} \ie have the form $\sumProc$. The entry place
of the subnet is connected to the entry places of the
translations of each $S_i$ by transitions. In case of
communication actions $\pi_i\neq\tau$, these transitions are
stubs that carry appropriate synchronisation actions and name
binding commands.\medskip

\noindent\textbf{Parallel composition and hiding}\quad Given
two stub transitions $t'$ and $t''$ in different threads
representing prefixes $\send{a}{b}$ and $\rec{x}{y}$, parallel
composition $\parallel$ adds a set of transitions implementing
the communication, \cf Fig.~\ref{fi-comm}. If static analysis
(see Sect.~\ref{se-opt}) shows that the prefixes are
potentially synchronisable, we create for each
$i\in\dom(a)\cap\dom(x)$ and $j\in\dom(b)\cap\dom(y)$ a
transition $t_{ij}$ which:
\begin{iteMize}{$\bullet$}
  \item consumes tokens from the input places of the stubs
      $t'$ and $t''$ and produces tokens on their output
      places;
  \item checks by read arcs that $[a=i]$ and $[x=i]$ are
      marked, \ie the substitution maps $a$ and $x$ to the
      same value $i$ and thus the synchronisation is
      possible (if $a$ and/or $x$ are in \PUB then the
      corresponding arc is not needed);
  \item checks by a read arc that $[b=j]$ is marked,
      consumes a token from $[y\neq j]$ (if this place
      exists) and produces a token on $[y=j]$ (mapping thus
      $y$ to the value of $b$).
\end{iteMize}
If the synchronisation is possible, exactly one of these
transitions is enabled (depending on the values of $a$, $b$,
and $x$); else none of these transitions is enabled. After all
such synchronisations are performed, the stub transitions are
removed from the net by hiding $H$.

\subsection{Example}\label{se-example}

\begin{wrapfigure}{R}{0.5\columnwidth}
  \centering
  \vspace*{-1.7em}
  \includegraphics[width=0.40\columnwidth,keepaspectratio]{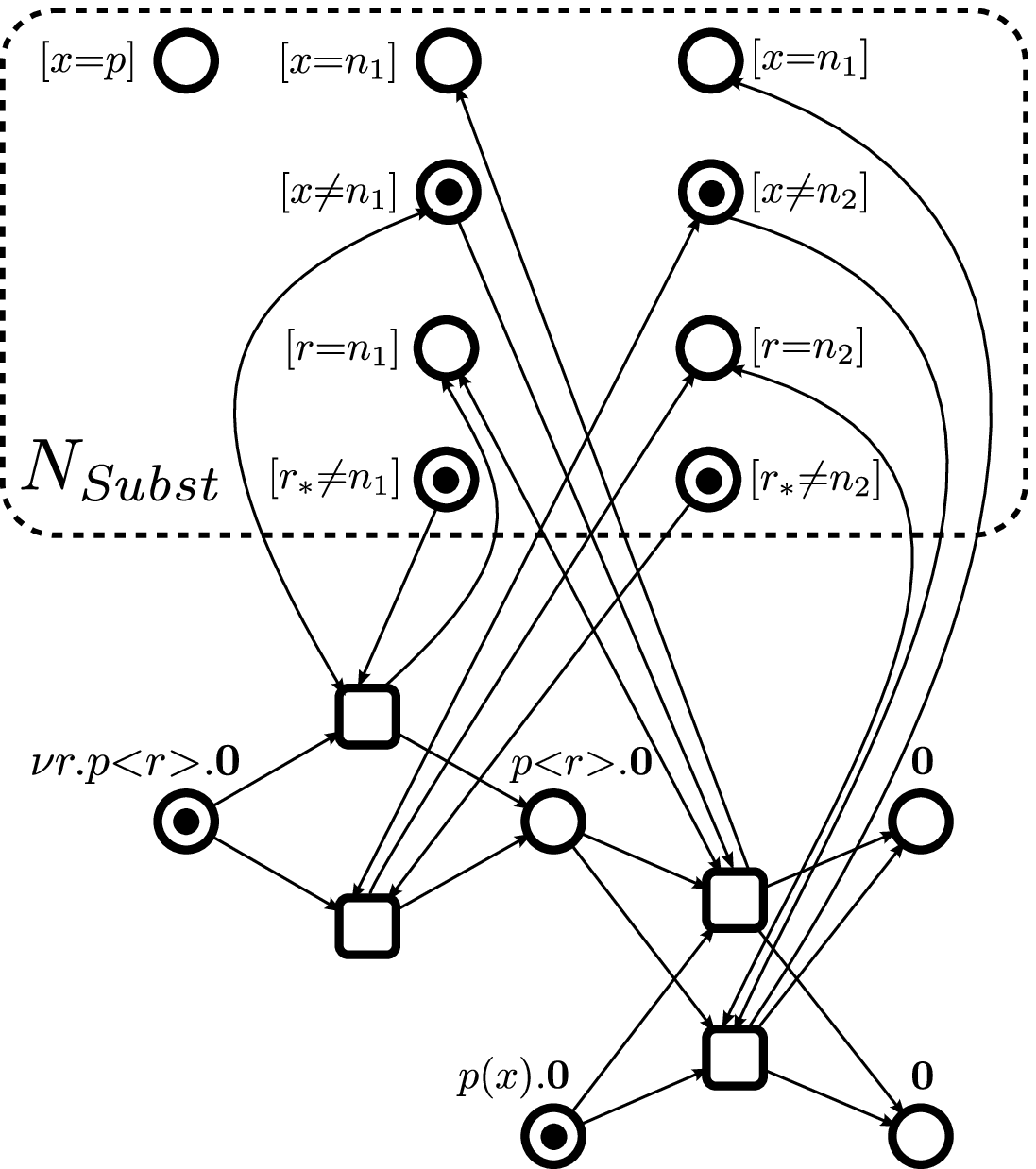}
  \caption{\label{fi-example}Translation of $\new r.\send{p}{r}.\stopProc\pc\rec{p}{x}.\stopProc$.}
  \vspace*{-2em}
\end{wrapfigure}

Fig.~\ref{fi-example} shows the complete translation of the FCP
$\new r.\send{p}{r}.\stopProc\pc\rec{p}{x}.\stopProc$. The
meanings of the places in $\nsubst$ are as in
Fig.~\ref{fi-subst}, and the places in the control flow are
labelled by the corresponding subterms.

As the FCP has two bound names, $r$ and $x$, we take
$\NEW=\{n_1,n_2\}$. The initialisation of $r$ is represented by
two transitions, corresponding to the values $n_1$ and $n_2$.
The only possible communication in this example is between the
prefixes $\send{p}{r}$ and $\rec{p}{x}$. Note that the
communication is over the public channel $p$, and the
communicated values are from $\dom(r)\cap\dom(x)=\{n_1,
n_2\}\cap\{p, n_1, n_2\}=\{n_1, n_2\}$. Hence there are two
transitions modelling this communication.

\section{Size of the resulting FCP}\label{se-size}
We now evaluate the contributions of various parts of the
translation to the size of the final safe PN. (Note that the
asymptotic size of a safe PN is fully determined by its total
number of places, transitions and arcs, as the size of the
initial marking is bounded by the number of places.) Recall
that we use $\NEW_{\RESTR\INP\FRM}$ as the set of names from
which the values for restricted names are picked.\medskip

\noindent\textbf{The substitution $\nsubst$}\quad This net
consists of
\begin{align*}
    (|\INP|+|\FRM|)\,|\PUB|
    +
    (2|\INP|+2|\FRM|+|\RESTR|+1)\,|\NEW_{\RESTR\INP\FRM}|
\end{align*}
places, with no transitions or arcs, see Fig.~\ref{fi-subst},
which is $O(\sizeApp{F}^2)$ in the worst case.\medskip

\noindent\textbf{Mapping a name in $\INP\cup\FRM$}\quad A
separate transition with $O(1)$ incident arcs is created for
each value in $\PUB\cup\NEW_{\RESTR\INP\FRM}$. Hence, the cost
of mapping a single input name or formal parameter is
$O(\sizeApp{F})$ in the worst case. (The cost of initialising a
restricted name is discussed later.)\medskip

\noindent\textbf{Unmapping a name in
$\INP\cup\FRM\cup\RESTR$}\quad A separate transition with
$O(1)$ incident arcs is created for each name in
$\PUB\cup\NEW_{\RESTR\INP\FRM}$ (in case of an input name or
formal parameter) or $\NEW_{\RESTR\INP\FRM}$ (in case of a
restricted name). Hence the cost of unmapping a single name is
$O(|\PUB|+|\NEW_{\RESTR\INP\FRM}|)=O(\sizeApp{F})$ in the worst
case.\medskip

\noindent\textbf{Stop processes}\quad A single place is created
for each occurrence of $\stopProc$, \ie the total contribution
is $O(\sizeApp{F})$ in the worst case.\medskip

\noindent\textbf{Calls}\quad In the worst case all the calls
have $O(\sizeApp{F})$ parameters in total, which have to be
mapped and unmapped. Hence the contribution is
$O(\sizeApp{F}^2)$.\medskip

\noindent\textbf{Restrictions}\quad In the worst case the
number of restrictions is $O(\sizeApp{F})$, and the
corresponding names have to be initialised and then unmapped.
For initialisation of a restricted name, a separate transition
is created for each value in $\NEW_{\RESTR\INP\FRM}$, and this
transition has $O(|\INP|+|\FRM|)$ incident arcs, see
Fig.~\ref{fi-restr}. Hence the contribution is
$O(\sizeApp{F}\,|\NEW_{\RESTR\INP\FRM}|\,(|\INP|+|\FRM|))=O(\sizeApp{F}^3)$.\medskip

\noindent\textbf{Choices and communication}\quad Implementing
the branching resulting from sums contributes $O(\sizeApp{F})$
to the final total. Furthermore, $\tau$-prefixes also
contribute at most $O(\sizeApp{F})$. In the worst case the
numbers of sending and receiving prefixes are $O(\sizeApp{F})$,
and almost all pairs of send/receive actions can synchronise;
thus the total number of such synchronisations is
$O(\sizeApp{F}^2)$. Recall that for a pair of actions
$\send{x_1}{y_1}$ and $\rec{x_2}{y_2}$, a separate transition
with $O(1)$ incident arcs is generated for each pair of values
in $\PUB\cup\NEW_{\RESTR\INP\FRM}$, see Fig.~\ref{fi-comm}.
Hence, the contribution is
$O(\sizeApp{F}^2\,|\PUB\cup\NEW_{\RESTR\INP\FRM}|^2)=O(\sizeApp{F}^4)$,
dominating thus the other parts of the translation. However,
the communication splitting optimisation described in
Sect.~\ref{se-opt} reduces this contribution down to
$O(\sizeApp{F}^3)$.\medskip

Totaling the above contributions shows that the size of the
resulting PN is $O(\sizeApp{F}^3)$. Furthermore, converting a
general FCP into a safe one can increase the size by a factor
bounded by the number of threads, \ie quadratically in the
worst case, see Sect.~\ref{se-basics}. Therefore, the
translation is polynomial for general FCPs too.

It should also be noted that the worst case size computed above
is rather pessimistic: the translation admits several
practical optimisations, see Sect.~\ref{se-opt}. The
experimental results in Sect.~\ref{se-experiments} demonstrate
that for realistic FCPs the sizes of the resulting PNs are
moderate.

\section{Definition of the translation}\label{Section:Translation}

We now formalise the proposed translation. To do that, we add
further assumptions on the form of the FCP. Again, these
assumptions are not restrictive: any FCP can be transformed
into the required form. However, the assumptions significantly
simplify the correctness proofs by reducing the number of cases
that have to be considered.

\subsection{Additional normal form assumptions}\label{se-additional-assumptions}
We augment the \noclash\ assumptions as follows: in a defining
equation $K(\tilde f):=S$, $\freeNames{S}= \tilde f$, \ie
public names are not allowed in~$S$. This assumption can be
enforced by passing the required public names as parameters.

To avoid case distinctions for the initial process, we assume there are artificial defining
equations $K_{\init, i}(\tilde f_{\init, i}):=S_{\init, i}$
with $\freeNames{S_{\init, i}}=\tilde f_{\init, i}\subseteq
\FRM$, that are called by a virtual initialisation step. Their
purpose is to guarantee that the $S_{\init, i}$ have
the free names $\tilde f_{\init, i}$. We then apply
substitutions to assign the expected values to these
parameters. This means we can write the given FCP as
\begin{align*}
F = \new \tilde r.(S_{\init, 1}\sigma_1\parComp\ldots \parComp
S_{\init, n}\sigma_n),
\end{align*}
where $\sigma_i:\tilde f_{\init, i}\rightarrow \tilde r\cup
\PUB$. We additionally assume that the $S_{\init, i}$ are
choices or calls and that the FCP does not contain the
$\stopProc$ process.

Moreover, if we have an input $\rec{x}{y}.S$ then we assume
$y\in\freeNames{S}$, which can be achieved by adding an
artificial parameter to the call at the end of the process.
Similarly, for a restriction $\new r.S$ we assume
$r\in\freeNames{S}$. Restrictions not satisfying this
requirement can be dropped due to structural congruence.

\subsection{Construction of $\nsubst$}
To represent a substitution like $\{a, b/x, y\}$, we decompose it into elementary substitutions
$\{a/x\}\cup\{b/y\}$ of single names.
The substitution net has corresponding places $\subeq{x}{a}$ and
$\subeq{y}{b}$ for each elementary substitution that may occur in such a decomposition.
Moreover, there is a second set of places, $\subneq{x}{n}$ and
$\subneq{r_*}{n}$, keeping track of whether an input, a formal
parameter, or a restriction is bound to $n\in \NEW$. These
places complement the corresponding substitution places, in
particular $\subneq{r_*}{n}$ indicates that no restricted name
is bound to $n$. (Since at most one restriction can be bound to
$n$, this one complement place is sufficient.)
$\nsubst$ has no transitions. We defer the explanation of its
initial marking for the moment.
Formally, $\nsubst\DEF(P_{\subst}\cup P_{\refcount}, \emptyset,
\emptyset, M_{0})$ with
\begin{align*}
P_{\subst}
    {\DEF}
    ((\INP\cup\FRM){\times}\{=\}{\times}\PUB)\cup ((\INP\cup\FRM\cup\RESTR){\times}\{=\}{\times}\NEW)
    \quad
    P_{\refcount}{\DEF}(\INP\cup\FRM\cup \{r_*\}){\times}\{\neq\}{\times}\NEW.
\end{align*}
\medskip

\noindent\textbf{Substitution markings and correspondence}\quad
A marking $M$ of $\nsubst$ is called a \emph{substitution
marking} if it satisfies the following constraints:
\begin{align*}
\substmarkrest\quad M(\subneq{r_*}{n})&+\sum_{r\in \RESTR}M(\subeq{r}{n})=
1\hspace{1cm}
\sum_{a\in\PUB\cup \NEW}M(\subeq{x}{a})\leq 1\quad \substmarkbind
\\
&M(\subeq{x}{n})+M(\subneq{x}{n})=1.\quad \substmarkcompl
\end{align*}
\substmarkrest{} holds for every $n\in\NEW$ and states that at
most one restricted name is bound to $n$. Moreover, there is a token
on $\subneq{r_*}{n}$ iff there is no such binding.
\substmarkbind{} states that every name
$x\in\INP\cup\FRM\cup\RESTR$ is bound to at most one
$a\in\PUB\cup \NEW$. The reference counter has to keep track of
whether a name $x\in\INP\cup \FRM$ maps to a fresh name
$n\in\NEW$, which motivates \substmarkcompl{}.

Consider now a substitution $\sigma:(\INP'\cup \FRM'\rightarrow
\PUB\cup\tilde r) \cup (\RESTR'\rightarrow \tilde r)$ where
$\INP'\subseteq \INP$, $\FRM'\subseteq \FRM$, $\RESTR'\subseteq
\RESTR$, and the second component $\RESTR'\rightarrow \tilde r$
is injective.
A substitution marking $M$ of $\nsubst$ is said to \emph{correspond to
$\sigma$} if the following hold:\vspace{0.2cm}
\begin{enumerate}[\hbox to8 pt{\hfill}]   
\item\noindent{\hskip-12 pt\cornotdom:}\ For all
    $x\in(\INP\cup\FRM\cup\RESTR)\setminus \dom(\sigma)$
    and $a\in\NEW\cup\PUB$,
    $M(\subeq{x}{a})=0$.\vspace{0.1cm}
\item\noindent{\hskip-12 pt\corpub:}\ For all $x\in \dom(\sigma)$ with
    $\sigma(x)\in\PUB$, $M(\subeq{x}{\sigma(x)})=1$.\vspace{0.1cm}
\item\noindent{\hskip-12 pt\correst:}\ For all $x\in \dom(\sigma)$ with
    $\sigma(x)\in\tilde r$, there is $n\in\NEW$ s.t.\
    $M(\subeq{x}{n})=1$.\vspace{0.1cm}
\item\noindent{\hskip-12 pt\coreq:}\ The choice of $n$ preserves the equality of
    names as required by $\sigma$: for all $x, y\in\dom(\sigma)$ with $\sigma(x),\sigma(y)\in\tilde r$, we have
\[
\sigma(x)=\sigma(y)\quad\text{iff}\quad
\mbox{for all $n\in \NEW$,~}M(\subeq{x}{n})=M(\subeq{y}{n}).
\]
\end{enumerate}

\noindent Recall that we translate the specification $F=\new \tilde
r.(S_{\init, 1}\sigma_1\parComp\ldots \parComp S_{\init,
n}\sigma_n)$. As the \emph{initial marking} of $\nsubst$, we
fix some substitution marking that corresponds to
$\sigma_1\cup\ldots \cup \sigma_n$\label{ref_initialmarkings}.
As we shall see, every choice of fresh names $\tilde n$ for
$\tilde r$ indeed yields bisimilar behaviour. Note that
\noclash\ ensures that the union of substitutions is again a
function. Fig.~\ref{fi-subst} illustrates $\nsubst$ and the
concepts of substitution markings and correspondence.


\subsection{Construction of $N(S_{\init})$}
\label{se-construction_ns}
Petri net $N(S_{\init})$ reflects the control flow of thread $S_{\init}$.
To synchronise send and receive prefixes in different threads,
we annotate its transitions with labels from
\begin{align*}
\mathcal{L}\DEF\set{\tau, \mathit{send}(a, b), \mathit{rec}(a,
b)}{a, b\in \PUB\cup \NEW}.
\end{align*}
To capture the effect that reactions have on substitutions, transitions also carry a set
of commands from
\begin{align*}
\mathcal{C}\DEF\set{\map{x}{b},
\unmap{x}{b}, \test{[x=b]}}{x\in\INP\cup\FRM\cup\RESTR\text{ and }b\in
\PUB\cup\NEW}.
\end{align*}

\begin{wrapfigure}{R}{0.46\columnwidth}
  \centering
  \includegraphics[width=0.42\columnwidth,keepaspectratio]{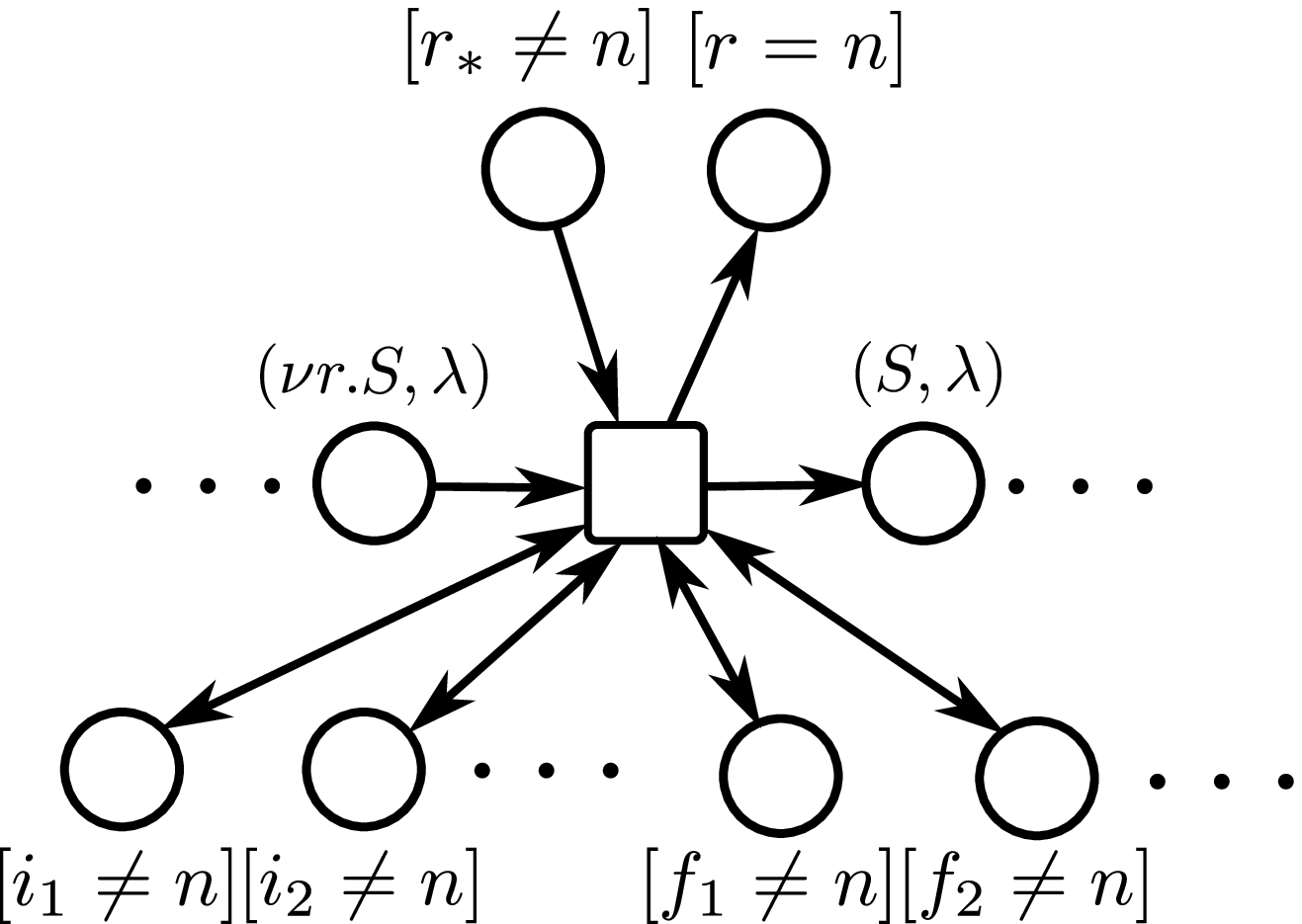}
  \caption{\label{fi-restr}Translation of a res\-t\-ric\-tion with
$\map{r}{n}$ im\-p\-le\-men\-ted.}
\end{wrapfigure}

\noindent Using these sets, a \emph{control flow net} is
defined to be a tuple $(P, T, F, M_0, l, c)$, where $(P, T, F,
M_0)$ is a PN and $l:T\rightarrow \mathcal{L}$ and
$c:T\rightarrow \powerSet{\mathcal{C}}$ are the transition
labelling functions.

As $S_{\init}$ is a sequential process, transitions in
$N(S_{\init})$ will always have a single input and a single
output place. This allows us to understand $N(S_{\init})$ as a
finite automaton, and hence define it implicitly via a new
labelled transition system for $S_{\init}$.
Recall that $\sProc$ is the set of sequential processes. We
augment them by lists of names, $\sProc\times
(\INP\cup\FRM\cup\RESTR)^*$, carrying the names that have been
forgotten and should be eventually unmapped in $\nsubst$.
Among such augmented processes, we define the labelled
transition relation
\begin{align*}
\pntrans{}{}\ \subseteq
(\sProc\times
(\INP\cup\FRM\cup\RESTR)^*)\times
\mathcal{L}\times\powerSet{\mathcal{C}}\times(\sProc\times
(\INP\cup\FRM\cup \RESTR)^*).
\end{align*}
Each transition carries a label and a set of commands, and will
yield a PN transition.

For restrictions $\new r.S$, we allocate a fresh name. Since we
can select any name that is not in use, such a transition
exists for every $n\in\NEW$:
\begin{align*}
(\new r.S,
\lambda)\pntrans{\{\map{r}{n}\}}{\tau}(S, \lambda) \tag{$\mathit{TRANS}_\nu$}\label{trans_new}.
\end{align*}
Fig.~\ref{fi-restr} depicts the transition, together with the
implementation of mapping defined below.

Silent actions yield $\tau$-labelled transitions with empty
sets of commands as expected:
\begin{align}
(\tau.S+\ldots, \lambda)\pntrans{\emptyset}{~~\tau~~}(S, \lambda\cdot\lambda'), \tag{$\mathit{TRANS}_\tau$}\label{trans_silent}
\end{align}
where $\lambda' =
\freeNames{\tau.S+\ldots}\setminus\freeNames{S}$ contains the
names that were free in the choice process but have been
forgotten in $S$. With an ordering on $\PUB\cup\NEW$, we can
understand this set as a sequence.

\begin{figure}[t]
  \centering
  \includegraphics[width=1.0\columnwidth,keepaspectratio]{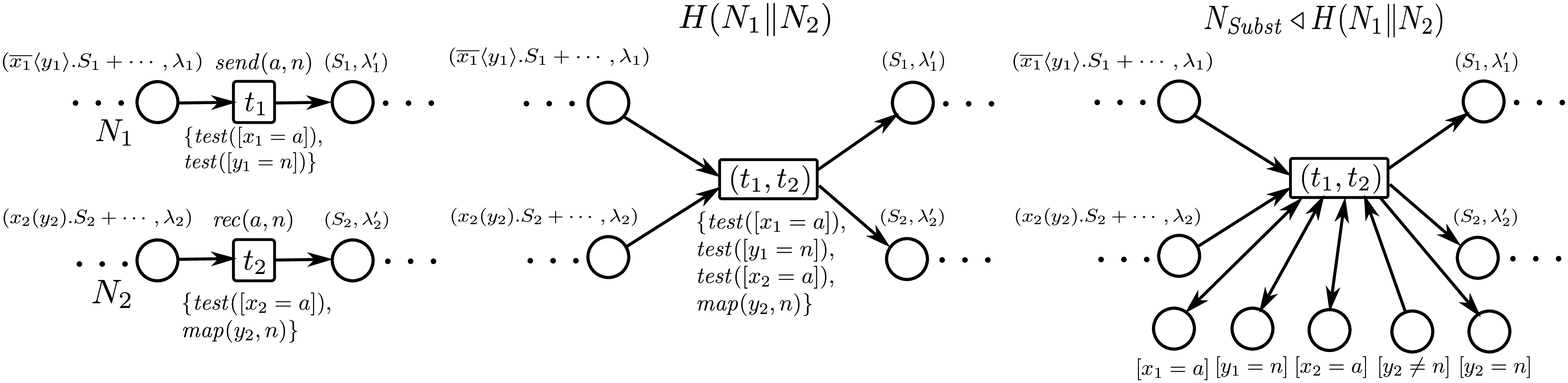}
  \caption{\label{fi-comm}Translation of communication (left), parallel
composition and hiding (center), and implementation of commands (right).}
\end{figure}

Communications are more subtle. Consider $\send{x}{y}.S+\ldots$
that sends $y$ on channel~$x$. With appropriate tests, we find
the names $a$ and $b$ to which $x$ and $y$ are mapped. These
names then determine the transition label. So for all $a, b\in
\PUB\cup \NEW$, we have
\begin{align*}
(\send{x}{y}.S+\ldots, \lambda)\pntrans{\{\test{\subeq{x}{a}},
\test{\subeq{y}{b}}\}}{\mathit{send}(a,
b)}(S, \lambda\cdot\lambda'). \tag{$\mathit{TRANS}_\mathit{snd}$}\label{trans_snd}
\end{align*}
Sequence $\lambda'$ again contains the names that have been
forgotten during this step. A receive action in
$\rec{x}{y}.S+\ldots$ is handled like a send, but introduces a
new binding. For all $a, b\in\PUB\cup \NEW$, we have
\begin{align*}
(\rec{x}{y}.S+\ldots, \lambda)\pntrans{\{\test{\subeq{x}{a}},
\map{y}{b}\}}{\mathit{rec}(a,
b)}(S, \lambda\cdot \lambda'). \tag{$\mathit{TRANS}_\mathit{rec}$}\label{trans_rec}
\end{align*}
There are similar transitions for the remaining
prefixes $\pi_i$ with $i\in I$. Fig.~\ref{fi-comm}(left)
illustrates the transitions for send and receive actions.

For a call $\call{K}{x_1,\ldots, x_{n}}$ with $K(f_1,\ldots,
f_{n}):=S$, the idea is to iteratively update the substitution
by binding the formal parameters to the factual ones. (Note
that we assumed $\freeNames{S}=\{f_1,\ldots,f_n\}$.)
Afterwards, we unmap the names in $\lambda$, which will then
include the factual parameters. Since no equation calls itself,
we do not accidentally unmap the just mapped formal parameters.
The following transitions are created for each
$a\in\PUB\cup\NEW$:
\begin{align*}
(\call{K}{x_1,\ldots, x_{m}},
\lambda)&\pntrans{\{\test{\subeq{x_m}{a}},
\map{f_m}{a}\}}{\tau}(\call{K}{x_1,\ldots, x_{m-1}},
\lambda'), \tag{$\mathit{TRANS}_{\mathit{call}_1}$}\label{trans_call1}
\end{align*}
where $\lambda'\DEF \lambda$ if $x_m\in\lambda$ and
$\lambda'\DEF\lambda\cdot x_m$ otherwise. (This case
distinction ensures that we will unmap a factual parameter
precisely once, even if it occurs multiple times in the list of
factual parameters.) When all parameters have been passed, we
unmap the names in $\lambda\neq\varepsilon$, by creating the
following transitions for each $a\in\PUB\cup\NEW$:
\begin{align*}
(\call{K}{-}, x\cdot \lambda)\pntrans{\{\unmap{x}{a}\}}{\tau}
(\call{K}{-}, \lambda). \tag{$\mathit{TRANS}_{\mathit{call}_2}$}\label{trans_call2}
\end{align*}
When $\lambda= \varepsilon$ has been reached, we transfer the control to the
body $S$ of the defining equation:
\[
    (\call{K}{-},\varepsilon)\pntrans{\emptyset}{~~\tau~~}(S, \varepsilon). \tag{$\mathit{TRANS}_{\mathit{call}_3}$}\label{trans_call3}
\]

Petri net $N(S_{\init})$ is the restriction of $(\sProc\times
(\INP\cup\FRM\cup\RESTR)^*, \pntrans{}{})$ to the augmented
processes that are reachable from $(S_{\init}, \varepsilon)$
via $\pntrans{}{}$. The initial marking puts one token on place
$(S_{\init}, \varepsilon)$ and leaves the remaining places
unmarked.

\subsection{Operations on nets} We now describe the operations
composing the nets.\medskip

\noindent\textbf{Parallel composition $\parallel$}\quad
Parallel composition of labelled nets is classical in Petri net
theory. The variant we use is inspired by
\cite{BestDevillersKoutny2001}: $N_1\parallel N_2$ forms the
disjoint union of $N_1$ and $N_2$, and then synchronises the
transitions $t_1$ in $N_1$ that are labelled by
$l_1(t_1)=\mathit{send}(a, b)$ (resp.\ $\mathit{rec}(a, b)$)
with the transitions $t_2$ in $N_2$ that are labelled by
$l_2(t_2)=\mathit{rec}(a, b)$ (resp.\ $\mathit{send}(a, b)$).
The result is a new transition $(t_1, t_2)$ labelled by $\tau$,
which carries the (disjoint) union of the commands for $t_1$
and $t_2$. Note that the transitions of $N_1$ and $N_2$ are
still available for further synchronisations with some $N_3$.
This in particular implies that $\parallel$ is associative and
commutative.\medskip

\noindent\textbf{Hiding $H$}\quad The \emph{hiding operator}
removes from a PN $N$ all transitions $t$ with $l(t)\neq\tau$.
Since $H(N)$ contains only $\tau$-labelled transitions, we can
omit the labelling function from the result. The combination of
parallel composition and hiding is illustrated in
Fig.~\ref{fi-comm}(center).\medskip

\noindent\textbf{Implementation operation $\impl$}\quad
Consider the two Petri nets $N_1=\nsubst=(P_1, \emptyset,
\emptyset, M_{0, 1})$ and $N_2=H(N(S_{\init, 1})\parallel
\ldots \parallel N(S_{\init, n}))=(P_2, T, F_2, M_{0, 2}, c)$
defined so far. The implementation operation
\begin{align*}
N_1\impl N_2 \DEF (P_1\cup P_2, T, F_2\cup F, M_{0, 1}\cup M_{0, 2})
\end{align*}
yields a standard Petri net without labelling. Its purpose is
to implement the commands labelling the transitions of $N_2$ by
adding arcs between the two nets. We fix a transition $t\in T$
and a command $c\in c(t)$, and define the arcs that have to be
added between $t$ and some places of $N_1$ to implement $c$. We
do the case analysis for the possible types of $c$:\medskip

\noindent $\test{\subeq{x}{b}}$\quad We add a loop to place
     $\subeq{x}{b}$: $(\subeq{x}{b}, t),(t,
     \subeq{x}{b})\in F$.\medskip

\noindent $\map{x}{p}, \map{x}{n}, \map{r}{n}$\quad A map
command
    differentiates according to whether the first component
    is an input name or a formal parameter $x\in\INP\cup \FRM$,
    or whether it is a restricted name $r\in\RESTR$. If $x$
    is assigned a public name, $\map{x}{p}\in c(t)$ with
    $p\in\PUB$, we add an arc producing a token in the
    corresponding place of the substitution
    net: $(t, \subeq{x}{p})\in F$. If $x$ is assigned some
    $n\in\NEW$, $\map{x}{n}\in c(t)$,
    we additionally remove the token from the reference
    counter: $(t,\subeq{x}{n}),(\subneq{x}{n}, t)\in F$.
To represent the restricted name $r\in\RESTR$ by a name
$n\in\NEW$, we first check that no other name is currently
mapped to $n$ using the reference counter for $n$. In case $n$
is currently not in use, we introduce the binding
$\subeq{r}{n}$ to the substitution net: $(\subneq{r_*}{n},
t),(t, \subeq{r}{n})\in F$ and $\{(\subneq{x}{n},
t),(t,\subneq{x}{n})\mid x\in\INP\cup\FRM\}\subseteq
F$.\medskip

\noindent $\unmap{x}{p}, \unmap{x}{n}, \unmap{r}{n}$\quad An
unmap
    removes the binding of $x\in\INP\cup\FRM$:
    $(\subeq{x}{p/n}, t)\in F$. Moreover, if
    $n\in\NEW$ it updates the reference counter:
    $(t,\subneq{x}{n})\in F$. When we remove the binding of $r\in\RESTR$ to $n\in\NEW$,
we update $\subneq{r_*}{n}$ in the reference counter:
$(\subeq{r}{n}, t),(t, \subneq{r_*}{n})\in F$.\medskip

Fig.~\ref{fi-restr} illustrates the implementation of mapping
for a restriction, $\map{r}{n}$. Tests and mapping of an input
name are shown in Fig.~\ref{fi-comm}(right).

\section{Correctness of the translation}

To prove the translation correct, we relate
$F$ and $N(F)$ by a suitable form of bisimulation.
%
The problem is that $N(F)$ may perform several steps to mimic one
transition of $F$.
The reason is that changes to substitutions (as induced \eg by
$\new r.S$) are handled by transitions in $N(F)$ whereas $F$
uses structural congruence, \ie a substitution change does not
necessarily lead to a step in the reaction relation of $F$.
To obtain a clean relationship between the models, we restrict
the transition system of $N(F)$ to so-called \emph{stable
markings} and \emph{race free transition sequences} between
them. Intuitively, stable markings correspond to the choices
and process calls in $F$, and race free transition sequences
mimic the reaction steps between them.
We show below that this restriction is insignificant, as any
transition sequence is equivalent to some race free one.

Place $(S, \lambda)$ of $N(F)=\nsubst\impl H(N(S_{\init,
1})\parallel\ldots\parallel N(S_{\init, n}))$ is called \emph{stable} if $S$
is a choice or a call to a process identifier with full parameter list.
Marking $M$ of $N(F)$ is called
\emph{stable} if, in every control flow net $N(S_{\init, i})$,
it marks a stable place.
We denote by $\reachstableof{N(F)}$ the set of stable markings
that are reachable in $N(F)$.

A transition sequence $t_1, \ldots, t_n$ between stable markings
$M, M'\in\reachstableof{N(F)}$ is \emph{race free} if
exactly one $t_i$ is either of the form \eqref{trans_silent}
for a silent action, of the form $(t,t')$ for communication
actions \eqref{trans_snd}, \eqref{trans_rec}, or of the form
\eqref{trans_call3} for an identifier call, \cf
Sect.~\ref{se-construction_ns}.
Thus, a race free transition sequence corresponds to precisely
one step in the reaction relation of $F$, characterised by
$t_i$, while the other transitions implement the substitution
changes between $M$ and $M'$. In particular, no intermediary
marking is stable. We denote the fact that there is such a race
free transition sequence by $M\Rightarrow M'$.

We now show that every transition sequence reaching a stable
marking $M$ can be replaced by a series of race free transition
sequences. This means the restriction to race free sequences is
inconsequential.

\begin{lem}\label{Lemma:reach_racefree}
For every transition sequence $M_1\rightarrow^+ M_2$ between
$M_1, M_2\in\reachstableof{N(F)}$, there is a sequence $M_1
\Rightarrow^+ M_2'$ with $M_2'\in\reachstableof{N(F)}$ and
\begin{iteMize}{$\bullet$}
  \item the control flow parts of $M_2$ and $M_2'$
      coincide, and
  \item the substitution parts of $M_2$ and $M_2'$
      correspond to a same substitution.
\end{iteMize}
Moreover, the latter sequence is a rearrangement of the former
one, modulo transitions implementing map and unmap operations
for restricted names using different values.
\end{lem}
\proof To obtain $M_1 \Rightarrow^+ M_2'$, we proceed as
follows. We project the sequence $M_1\rightarrow^+ M_2$ to the
transitions that reflect \picalc\ reactions. After each such
transition, we insert the required initialisations of
restricted names by unused values (not necessarily the same as
in the original sequence). Before each call to a process
identifier, we insert the necessary unmapping operations (the
ones for restricted names are amended to use the values given
during the corresponding initialisations). The result is race
free. To see that the sequence is enabled, note that
initialisations of restricted names cannot be blocked because
the pool of fresh names is large enough. Moreover, unmap
operations can never be blocked.
\qed

\subsection{Bisimulation}

Since the initial marking $M_0$ of $N(F)$ is stable by the
assumption on $S_{\init, i}$ from
Sect.~\ref{se-additional-assumptions}, we can define the
\emph{stable transition system of $N(F)$} as
\begin{align*}
\tsstableof{N(F)}\DEF(\reachstableof{N(F)}, \Rightarrow, M_{0}).
\end{align*}
\begin{thm}\label{Theorem:Bisimulation}
The transition system of $F$ and the stable transition system of $N(F)$ are
bisimilar, $\tsof{F}\sim \tsstableof{N(F)}$, via the bisimulation $\rel$
defined below.
\end{thm}
We defer the proof for the moment.
To define the bisimulation relation, we use the fact that every
process reachable from $F$ is structurally congruent to some
$\new \tilde r.(S_1\sigma_1\parComp\ldots \parComp
S_n\sigma_n)$. Here, $S_i$ is a choice or an identifier call
that has been derived from some process $S$ with $K(\tilde
f):=S$.
Derived means $(S, \varepsilon)\pntrans{}{}^+ (S_i,
\lambda_i)$ so that no intermediary process is a call to a
process identifier.
As second requirement, we have
\begin{align*}
\sigma_i: \freeNamesApp{S_i}\cup \lambda_i\rightarrow \tilde
r\cup\PUB.\quad\domproc
\end{align*}
This means the domain of $\sigma_i$ are the free names in $S_i$
together with the names $\lambda_i$ that have already been
forgotten. The two sets are disjoint, $\freeNamesApp{S_i}\cap
\lambda_i=\emptyset$.
%
The above process actually is in Milner's standard form \cite{Milner1999}, but
makes additional assumptions about the shape of threads and the domain of
substitutions.
%

We define $\rel \subseteq
\factorize{\reachof{F}}{\structCong}\times
\reachstableof{N(F)}$ to contain $(\class{G}, M_1\cup M_2)\in
\rel$ if there is a process $\new \tilde
r.(S_1\sigma_1\parComp\ldots \parComp S_n\sigma_n)\structCong
G$ as above so that the following hold:
\begin{iteMize}{$\bullet$}
  \item marking $M_1$ of $\nsubst$ corresponds to
      $\sigma_1\cup\ldots \cup \sigma_n$ and
  \item for the control flow marking, we have $M_2(S_i,
      \lambda_i)=1$ for all $i\in\{1,\ldots, n\}$.
\end{iteMize}

To relate $\tsof{F}$ and the full transition system
$\tsof{N(F)}$, consider a transition sequence $M_1\rightarrow^+
M_2$ between stable markings $M_1, M_2\in \reachstableof{N(F)}$
that need not be race free. Due to
Lemma~\ref{Lemma:reach_racefree} it can be rearranged to a race
free sequence $M_1\Rightarrow^+ M_2'$. With the above
bisimilarity, this race free transition sequence is mimicked by
a sequence of \picalc{} transitions $\class{F}\rightarrow^+
\class{G}$ with $(\class{G}, M_2')\in \rel$. With
Lemma~\ref{Lemma:reach_racefree} and the definition of $\rel$,
we also have $(\class{G}, M_2)\in\rel$. In the reverse
direction, a single process transition is still mimicked by a
sequence of PN transitions (that happens to be race free).
Hence, the following holds:
\begin{thm}\label{Theorem:WeakBisimulation}\hspace*{-1ex}
The transition systems of $F$ and $N(F)$ are weakly bisimilar,
$\tsof{F}{\approx}\tsof{N(F)}$, taking $\rel$ defined above as
a weak bisimulation.
\end{thm}
\noindent This result allows one to check temporal properties
of FCPs using their PN representations.

\subsection{Bisimulation Proof}
\label{se-bisim-proof}

We now turn to the proof of Theorem~\ref{Theorem:Bisimulation}.
We have to show that for each pair $(\class{G}, M)\in\rel$,
every transition $\class{G}\rightarrowts \class{G'}$ can be
mimicked by a race free transition sequence in $N(F)$, \ie
there is a stable marking $M'$ with $M \Rightarrow M'$ such
that $(\class{G'}, M')\in \rel$. Moreover and in turn, the race
free transition sequences in $N(F)$ should be imitated in
process $F$.
The proof is split into two parts, formulated as
Lemmas~\ref{Lemma:ProcToPN} and~\ref{Lemma:PNToProc}, for both
directions respectively.
\begin{lem}\label{Lemma:ProcToPN}
Consider $(\class{G}, M)\in\rel$. For all $\class{G'}$ with
$\class{G}\rightarrowts \class{G'}$ there is a stable marking
$M'\in \reachstableof{N(F)}$ such that $M\Rightarrow M'$ and
$(\class{G'}, M')\in \rel$.
\end{lem}
\proof Process $G$ is structurally congruent to $\new \tilde
r.(S_1\sigma_1\parComp\ldots \parComp S_n\sigma_n)$. By the
base cases of the reaction rules, transition
$\class{G}\rightarrowts \class{G'}$ exists iff (1) either two
processes $S_i\sigma_i$ and $S_j\sigma_j$ with $i\neq
j\in\{1,\ldots, n\}$ communicate, (2) we resolve a call to a
process identifier in some $S_i\sigma_i$, $i\in\{1,\ldots,
n\}$, or (3) we have a $\tau$ action. Silent steps are easier
than the former two and hence omitted in the proof.\medskip

\noindent\textbf{Case 1: Communication}\quad For simplicity, we
assume that: the first two threads communicate using the first
prefixes; after the communication, the first thread yields
choice or call $S_1'$; the second process creates precisely one
restricted name before becoming a choice or a call $S_2'$; the
communication is over restricted names and a restricted name is
sent. The remaining cases are along similar lines. We thus have
$G\structCong \new \tilde r.(S_1\sigma_1\parComp\ldots \parComp
S_n\sigma_n)$ with
\[
    S_1 = \send{x_1}{y_1}.S_1' + \ldots
    \quad
    S_2 = \rec{x_2}{y_2}.\new r.S_2' + \ldots
    \quad
    \sigma_1(x_1)=\sigma_2(x_2)\in \tilde r
    \quad
    \sigma_1(y_1)\in\tilde r.
\]
The process resulting from the communication is
\begin{align*}
G'\DEF \new \tilde r.a_r.(S_1'\sigma_1\parComp
S_2'\sigma_2'\parComp S_3\sigma_3\parComp\ldots
\parComp S_n\sigma_n)\quad\text{with}\quad \sigma_2' \DEF
\sigma_2\{\sigma_1(y_1)/y_2\}\{a_r/r\}.
\end{align*}
We argue that $G'$ has the desired normal form.
The processes $S_1'$ and $S_2'$ are choices or calls.
Moreover, $(S, \varepsilon)\pntrans{}{}^* (S_2, \lambda_2)$ implies
$(S, \varepsilon)\pntrans{}{}^* (S_2', \lambda_2\cdot\lambda_2')$.
This means $S_1'$ and $S_2'$ have been derived, as required.
It remains to show \domproc. We do the proof for $\sigma_2'$, the reasoning for
$\sigma_1$ is simpler:
\begin{align}
&\phantom{\text{= }}\ \dom(\sigma_2')\\
&=\dom(\sigma_2)\cup\{y_2, r\}\\
&=\lambda_2\cup
\freeNames{S_2}\cup\{y_2, r\}\label{Eq:DOMBISIM}\\
&=\lambda_2 \cup
(\freeNames{S_2}\setminus \freeNames{\new r.S_2'})\cup
(\freeNames{\new r.S_2'}\setminus\{y_2\})\cup\{y_2, r\}\label{Eq:Trick}\\
&=\lambda_2 \cup
(\freeNames{S_2}\setminus \freeNames{\new r.S_2'})\cup
\freeNames{S_2'}\cup\{y_2, r\}\label{Eq:Prime}\\
&=\lambda_2 \cup
(\freeNames{S_2}\setminus \freeNames{\new r.S_2'})\cup
\freeNames{S_2'}\label{Eq:In}\\
&=\lambda_2\cdot \lambda_2'\cup \freeNames{S_2'}\label{Eq:Def}.
\end{align}
Equation~\eqref{Eq:DOMBISIM} is \domproc\ for $\sigma_2$.
Equation~\eqref{Eq:Trick} uses the fact that
\begin{align*}
\freeNames{S_2} = (\freeNames{S_2}\setminus \freeNames{\new r.S_2'})\cup
(\freeNames{\new r.S_2'}\setminus\{y_2\}).
\end{align*}
This is due to $\freeNames{\new
r.S_2'}\setminus\{y_2\}\subseteq \freeNamesApp{S_2}$.
Equation~\eqref{Eq:Prime} is due to
\begin{align*}
(\freeNames{\new r.S_2'}\setminus\{y_2\})\cup\{y_2, r\} =
\freeNames{S_2'}\cup\{y_2, r\}.
\end{align*}
Equation~\eqref{Eq:In} holds by $\{y_2, r\}\subseteq
\freeNamesApp{S_2'}$. Finally, Equation~\eqref{Eq:Def} holds by
definition of the augmented transition relation
$\pntrans{}{}$.\medskip

We now argue that (1.a) there is $M'\in\reachstableof{N(F)}$ so
that $M\Rightarrow M'$ and (1.b) $(\class{G'},
M')\in\rel$.\medskip

\noindent\textbf{Claim 1.a: There is
$M'\in\reachstableof{N(F)}$ with $M\Rightarrow M'$}\quad Let
$M=M_1\cup M_2$ so that $M_1$ is the substitution marking and
$M_2$ is the control flow marking. Since $(\class{G}, M_1\cup
M_2)\in\rel$, we have $M_2((S_1, \lambda_1))=1=M_2((S_2,
\lambda_2))$. Moreover, $M_1$ corresponds to
$\sigma_1\cup\ldots \cup \sigma_n$. In the following, we also
use $\sigma$ to refer to this union. Since $\sigma_1(x_1),
\sigma_2(x_2), \sigma_1(y_1)\in\tilde r$, by \correst\ we have
fresh names $n_1, n_2, n_3\in\NEW$ with
$M_1(\subeq{x_1}{n_1})=1=M_1(\subeq{x_2}{n_2})=M_1(\subeq{y_1}{n_3})$.
Since $\sigma_1(x_1)=\sigma_2(x_2)$, we conclude $n_1=n_2$ by
\coreq.

It remains to show that there is a fresh name available in
$\NEW$ which we can use to represent $r$. As $r\notin\dom(\sigma)$, we have
\begin{align*}
\power{\dom(\sigma)}<\power{\INP}+\power{\FRM}
+\power{\RESTR}=\power{\NEW}.
\end{align*}
With \cornotdom, for $x\notin\dom(\sigma)$ we have
$M_1(\subeq{x}{n})=0$ for all $n\in\NEW$. For
$x\in\dom(\sigma)$, we have at most one place $\subeq{x}{a}$
marked by $\substmarkbind$. Together, these mean there is a
name $n\in\NEW$ with $M_1(\subeq{x}{n})=0$ for all
$x\in\INP\cup\FRM\cup\RESTR$. Let this name be $n_{r}$. As
$M_1(\subeq{x}{n_{r}})=0$ for $x\in\INP\cup\FRM\cup\RESTR$,
\substmarkrest\ and \substmarkcompl\ ensure
$M_1(\subneq{x}{n_{r}})=1$ for $x\in \INP\cup\FRM\cup\{r_*\}$.

Before parallel composition, the original net $N(S_{\init, 1})$ had the
following transition sequence leaving place $(S_1, \lambda_1)$:
\begin{align*}
(S_1, \lambda_1)\pntrans{\{\test{\subeq{x_1}{n_1}},
\test{\subeq{y_1}{n_3}}\}}{\mathit{send}(n_1, n_3)}(S_1',
\lambda_1').
\end{align*}
Similarly, from $(S_2, \lambda_2)$ in $N(S_{\init, 2})$ we get
\begin{align*}
(S_2, \lambda_2)\pntrans{\{\test{\subeq{x_2}{n_1}},
\map{\subeq{y_2}{n_3}}\}}{\mathit{rec}(n_1, n_3)}(\new r.S_2',
\lambda_2\cdot\lambda_2')\pntrans{\{\map{r}{n_{r}}\}}{\tau} (S_2',
\lambda_2\cdot\lambda_2').
\end{align*}
Parallel composition joins the communicating transitions of the
two nets, and we denote the result by $(t_1, t_2)$. Then hiding
removes the original transitions $t_1$ labelled by
$\mathit{send}(n_1, n_3)$ and $t_2$ labelled by
$\mathit{rec}(n_1, n_3)$. Then, for $(t_1, t_2)$ and for the
transition $t_r$ mapping $r$ to a fresh name, the
implementation operation adds arcs to and from $\nsubst$.

We now show that the transition sequence $(t_1, t_2)\, t_{r}$
is enabled. We argued
that $(S_i, \lambda_i)$ carries a token. This means the control
flow is at the right place. We have
$M_1(\subeq{x_1}{n_1})=1=M_1(\subeq{x_2}{n_1})=M_1(\subeq{y_1}{n_3})$.
Hence, the test arcs to the substitution net are enabled. We
have $y_2\in \boundNamesApp{S_2}$. Hence, the name is not in
the domain of $\sigma_2$ by \domproc\ and \noclash. With
\cornotdom, $M(\subeq{y_2}{a})=0$ holds for all names
$a\in\PUB\cup \NEW$. In particular, $M(\subeq{y_2}{n_3})=0$.
With \substmarkcompl, we conclude $M(\subneq{y_2}{n_3})=1$.
This ensures $\map{y_2}{n_3}$ is enabled. For $t_{r}$, we have
$M(\subneq{x}{n_{r}})=1$ for all $x\in
\INP\cup\FRM\cup\{r_*\}$. Hence, the transition is enabled.

The resulting marking $M'$ puts tokens on $(S_1', \lambda_1')$
and $(S_2', \lambda_2\cdot \lambda_2')$ which are stable
places. This means $M'$ is stable. The marking is reachable as
$M$ was reachable. Moreover, transition sequence $(t_1,
t_2)\cdot t_r$ above is race free.\medskip

\noindent\textbf{Claim 1.b: $(\class{G'}, M')\in\rel$}\quad
Again $M' = M_1'\cup M_2'$ where $M_1'$ is the marking of
$\nsubst$ and $M_2'$ is the control flow. For the control flow,
we moved the single token from $(S_1, \lambda_1)$ to $(S_1',
\lambda_1')$ and from $(S_2, \lambda_2)$ to $(S_2',
\lambda_2\cdot \lambda_2')$ as required.

For $\nsubst$, we show that we obtain a substitution marking.
We already argued that $M_1(\subeq{y_2}{a})=0$ for all
$a\in\PUB\cup \NEW$ and hence $M_1(\subneq{y_2}{n_3})=1$. We
consume the latter token and move it to
$M_1'(\subeq{y_2}{n_3})=1$. This means we still map $y_2$ to at
most one name as required by \substmarkbind. Moreover, the
invariant on reference counting \substmarkcompl\ is satisfied.

Name $r$ is not in the domain of $\sigma_2$.
Hence, the places $\subeq{r}{a}$ are empty for all $a\in\NEW\cup \PUB$.
We move the token from $M_1(\subneq{r_*}{n_{r}})=1$ to
$M_1'(\subeq{r}{n_r})=1$.
As a result, the places $\subeq{r}{a}$ for all $a\in\NEW\cup\PUB$ together carry
at most one token as required by $\substmarkbind$.
Moreover, the places $\subeq{r}{n_{r}}$ for all $r\in\RESTR$ plus
$\subneq{r_*}{n_{r}}$ carry
precisely one token.
This proves $\substmarkrest$.
We have a substitution marking.

We have to show that $M_1'$ corresponds to
$\sigma'\DEF\sigma_1\cup \sigma_2'\cup\sigma_3\cup\ldots \cup
\sigma_n$. We only introduce bindings for $y_2$ and $r$. For
$y_2$ we have $\sigma_2'(y_2)=\sigma_1(y_1)\in\tilde r$. Hence,
it is correct that we map $M_1'(\subeq{y_2}{n_3})=1$ with
$n_3\in\NEW$. The reasoning is similar for $r$ with
$\sigma_2'(r)=a_r$. \correst\ holds. Marking $M_1'$ only
introduces tokens to the places $\subeq{y_2}{n_3}$ and
$\subeq{r}{n_{r}}$ with $\{y_2, r\}\subseteq \dom(\sigma')$.
For the remaining names
$x\in\INP\cup\FRM\cup\RESTR\setminus\{y_2, r\}$, it coincides
with $M_1$. Note that for $x\notin\dom(\sigma')$ we have
$x\notin \dom(\sigma)$. Hence, by \cornotdom\ for $M_1$, we get
$M_1'(\subeq{x}{a})=M_1(\subeq{x}{a})=0$ for all
$a\in\PUB\cup\NEW$. This proves \cornotdom\ for $M_1'$.

It remains to show \coreq: the equality required by $\sigma'$ coincides
with the
choice of fresh names.
For $r$ we have $M_1'(\subeq{r}{n_{r}})=1$ and $M_1'(\subeq{x}{n_{r}})=0$ for
all other names $r\neq x\in\INP\cup\FRM\cup\RESTR$.
This coincides with the requirement that $\sigma'(r)\neq \sigma'(x)$.
For $y_2$, we only consider $x\notin\{y_2, r\}$ and get
\begin{align*}
\sigma'(y_2) = \sigma'(x)\quad &\text{iff}\quad \sigma(y_1) = \sigma(x)\\
&\text{iff } M_1(\subeq{y_1}{n}) = M_1(\subeq{x}{n})\text{ for all }n\in\NEW\\
&\text{iff } M_1'(\subeq{y_1}{n}) = M_1'(\subeq{x}{n})\text{ for all }n\in\NEW\\
&\text{iff } M_1'(\subeq{y_2}{n}) = M_1'(\subeq{x}{n})\text{ for all }n\in\NEW.
\end{align*}
The first equivalence holds by $\sigma'(y_2)=\sigma(y_1)$. The
second equivalence is \coreq\ for $\sigma$, the third is the
observation that $M_1$ and $M_1'$ coincide on all names except
$y_2$ and $r$. The last equivalence is the fact that the rows
for $y_1$ and $y_2$ coincide. This is by \substmarkbind, in
combination with
$M_1'(\subeq{y_1}{n_3})=1=M_1'(\subeq{y_2}{n_3})$.\medskip

\noindent\textbf{Case 2: Identifier calls}\quad We have
$G\structCong \new \tilde r.(\call{K}{\tilde
x}\sigma_1\parComp\ldots \parComp S_n\sigma_n)$ with $K(\tilde
f):=S$. We assume $S$ already is a choice or a call. The
process resulting from the call $\call{K}{\tilde x}\sigma_1$ is
\begin{align*}
G'\DEF \new \tilde r.(S\sigma_1'\parComp \ldots
\parComp S_n\sigma_n)\quad\text{with}\quad \sigma_1' \DEF
\{\sigma_1(\tilde x)/\tilde f\}.
\end{align*}
We argue that $G'$ has the desired normal form.
The process $S$ is a choice or a call. It has been derived trivially as it is
the defining process.
For \domproc, we have as desired
\begin{align*}
\dom(\sigma_1')=\tilde
f=\freeNames{S}=\freeNames{S}\cup\emptyset.
\end{align*}
We now argue that (2.a) there is $M'\in\reachstableof{N(F)}$ so
that $M\Rightarrow M'$ and (2.b) $(\class{G'},
M')\in\rel$.\medskip

\noindent\textbf{Claim 2.a: There is
$M'\in\reachstableof{N(F)}$ with $M\Rightarrow M'$}\quad Let
$M=M_1\cup M_2$ so that $M_1$ is the substitution marking and
$M_2$ is the control flow marking. Since $(\class{G}, M_1\cup
M_2)\in\rel$, we know that $M_1$ corresponds to
$\sigma\DEF\sigma_1\cup\ldots \cup \sigma_n$. For $M_2$, we
have $M_2(\call{K}{\tilde x}, \lambda)=1$. Moreover, by
\domproc, we have $\tilde x\cup \lambda = \dom(\sigma_1)$.
Hence, for every name $x_i\in\tilde x\cup\lambda$ we have a
name $a_i\in\PUB\cup\NEW$ so that $M_1(\subeq{x_i}{a_i})=1$ by
\corpub\ and \correst. Since an equation does not call itself
and since all formal parameters are unique by \noclash, we have
$\tilde f\cap \dom(\sigma)=\emptyset$ by \domproc. This means
$M_1(\subeq{f}{a})=0$ for all $f\in \tilde f$ and all $a\in
\NEW\cup\PUB$. With \substmarkcompl, we get
$M_1(\subneq{f}{n})=1$ for all $f\in \tilde f$ and all $n\in
\NEW$.

By definition, Petri net $N(S_{\init, 1})$ has the following transition
sequence:
\begin{align*}
(\call{K}{\tilde x}, \lambda)\pntrans{\{\test{\subeq{x_i}{a_{i}}},
\map{f_i}{a_i}\}}{\tau}^+(\call{K}{-},
\lambda')\pntrans{\{\unmap{x_i}{a_i}\}}{\tau}^+(\call{K}{-},
\varepsilon)\pntrans{\emptyset}{\tau}(S, \varepsilon).
\end{align*}
The first transition sequence introduces the bindings for
$\tilde f$ and moves the names in $\tilde x$ to $\lambda$. The
result is $(\call{K}{-}, \lambda')$ with $\lambda'=\lambda\cdot
\tilde x'$, where $\tilde x'$ is obtained from $\tilde x$ by
removing the duplicates. The next transition sequence unmaps
all names in $\lambda'$. Finally, the token is moved to $(S,
\varepsilon)$.

We now show that the composed sequence is enabled. For the first sequence, the
tests are enabled with $M_1(\subeq{x_i}{a_i})=1$.
For formal parameters, mapping $\map{f}{p}$ with $p\in\PUB$ is always
enabled, and $\map{f}{n}$ with $n\in\NEW$ requires
$M_1(\subneq{f}{n})=1$.
This holds by the above argumentation.
The second transition sequence removes the tokens from $\subeq{x_i}{a_i}$.
Since we do not repeat names in $\tilde x'$ and since $\tilde x\cap
\lambda =\emptyset$, all transitions are enabled.
For $a_i=n\in\NEW$, unmapping introduces a token to $\subneq{x_i}{n}$ or to
$\subneq{r_*}{n}$.

The resulting marking $M'$ puts a token on the stable place
$(S, \varepsilon)$, \ie $M'$ is stable. $M'$ is reachable as
$M$ was reachable. Moreover, the transition sequence above is
race free.\medskip

\noindent\textbf{Claim 2.b: $(\class{G'}, M')\in\rel$}\quad
Again we have $M' = M_1'\cup M_2'$ where $M_1'$ is the marking
of $\nsubst$ and $M_2'$ is the control flow marking. For the
control flow, we moved the single token from $(\call{K}{\tilde
x}, \lambda)$ to $(S, \varepsilon)$ as required.

For $\nsubst$, we show that we obtain a substitution marking.
We already argued that $M_1(\subeq{f}{a})=0$ for all
$a\in\PUB\cup \NEW$ and hence $M_1(\subneq{f}{n})=1$ for all
$n\in\NEW$. We introduce a token $M_1'(\subeq{f_i}{a_i})=1$,
potentially consuming the complement marking if $a_i=n\in\NEW$.
\substmarkbind\ holds: names are bound at most
once. The second transition sequence manipulates the places for
$\tilde x'\cup\lambda$. These names are disjoint from $\tilde
f$ due to $\tilde f\cap\dom(\sigma_1)=\emptyset$ explained
above. We remove all tokens
$M_1(\subeq{x_i}{a_i})=1$ with $x_i\in\tilde x'\cup\lambda$.
The implementation of unmap ensures we reinstall complement
markings. More precisely, if $x_i=r\in\RESTR$ and $a_i=n$, we
mark $M_1'(\subneq{r_*}{n})=1$. Since by \substmarkrest, name
$r$ was the only restriction bound to $n$, the constraint
continues to hold with $\subneq{r_*}{n}$ marked. If
$M_1(\subeq{x_i}{n})=1$ with $x_i\in\INP\cup\FRM$, we get
$M_1'(\subneq{x_i}{n})=1$. Hence, \substmarkcompl\ continues to
hold. We have a substitution marking.

We have to show that $M_1'$ corresponds to
$\sigma'\DEF\sigma_1'\cup \sigma_2\cup\ldots \cup \sigma_n$. We
focus on $\sigma_1'$ and assume $\sigma_1'(f_i)\in \tilde r$.
This means $\sigma_1(x_i)\in\tilde r$ for the corresponding
name $x_i\in \tilde x$. Since $M_1$ corresponds to $\sigma$, by
\correst\ for $M_1$ we have $M_1(\subeq{x_i}{n})=1$ for a name
$n\in \NEW$. By \substmarkbind, $x_i$ is bound to only one
name. This means $n$ has to be the name $a_i$, $n=a_i$, that we
chose for the transition. As a result, we have
$M_1'(\subeq{f}{n})=1$ with $n\in\NEW$ as required. For
$\sigma_1'(f)\in\PUB$, the reasoning is similar. For the names
in $\INP\cup\FRM\cup\RESTR\setminus(\dom(\sigma_1)\cup\tilde
f)$, markings $M_1$ and $M_1'$ coincide. Hence, if
$x\notin\dom(\sigma')$ we either have $x\notin \dom(\sigma)$ or
we have $x\in\dom(\sigma_1)$. In the former case, we get
$M_1'(\subeq{x}{a})=M_1(\subeq{x}{a})=0$ for all
$a\in\PUB\cup\NEW$ by \cornotdom\ for $\sigma$. In the latter
case, the name has been explicitly unmapped by the second
transition sequence. Hence, \cornotdom\ holds for $\sigma'$.

It remains to show \coreq: the equality required by $\sigma'$ coincides
with the choice of fresh names.
Consider $f_i, f_j\in\tilde f$:
\begin{align*}
\sigma'(f_i) = \sigma'(f_j)\quad &\text{iff}\quad \sigma(x_i) =
\sigma(x_j)\\
&\text{iff } M_1(\subeq{x_i}{n}) = M_1(\subeq{x_j}{n})\text{ for all }n\in\NEW\\
&\text{iff } M_1'(\subeq{f_i}{n}) = M_1'(\subeq{f_j}{n})\text{ for all
}n\in\NEW.
\end{align*}
The first equivalence holds by $\sigma_1'(f_i)=\sigma_1(x_i)$
and $\sigma_1'(f_j)=\sigma_1(x_j)$. The second is \coreq\ for $\sigma$. The
third equivalence is the fact that the rows for $x_i$ in $M_1$
and for $f_i$ in $M_1'$ coincide. This is by the fact that
$x_i$ and $f_i$ mark at most one place $\subeq{x_i}{a_i}$ and
$\subeq{f_i}{a_i}$ by \substmarkbind, and by the fact that this
name $a_i$ coincides. The reasoning for $\sigma'(f)=\sigma'(x)$
with $x\in\dom(\sigma_2\cup\ldots\cup \sigma_n)$ is
similar.\qed

We now turn to the reverse direction and argue that $\class{G}$
can imitate race free transition sequences enabled by $M$.
\begin{lem}\label{Lemma:PNToProc}
Let $(\class{G}, M)\in \rel$. For all $M'\in\reachstableof{N(F)}$ so that
$M\Rightarrow M'$ there
is a process $\class{G'}$ with $\class{G}\rightarrowts \class{G'}$ and
$(\class{G'},
M')\in\rel$.
\end{lem}
\proof A race free transition sequence $M\Rightarrow M'$
corresponds to a communication among two processes (1), to an
identifier call (2), or to a silent action (3). We only
consider the first case, the remaining two are along similar
lines.\medskip

\noindent\textbf{Case 1: Communication}\quad We reconstruct the
race free transition sequence $M\Rightarrow M'$ to derive
information about the shape of $M$ and $M'$. Since we model a
communication, we have $M(S_1, \lambda_1)=1$ in the net
$N(S_{\init, 1})$ with $S_1 = \send{x_1}{y_1}.S_1'+\ldots$.
Similarly, $M(S_2, \lambda_2)=1$ in $N(S_{\init, 2})$ with $S_2
= \rec{x_2}{y_2}.\new r.S_2'+\ldots$. Here, $S_1'$ and $S_2'$
are meant to be choices or identifier calls. Thus, again the
first two processes communicate and the second generates a
fresh name. The race free transition sequence $M\rightarrow^+
M'$ in $N(F)$ is now $(t_1, t_2)\, t_r$ where
\begin{align*}
t_1 &= (S_1, \lambda_1)\pntrans{\{\test{\subeq{x_1}{n_1}},
\test{\subeq{y_1}{n_2}}\}}{\mathit{send}(n_1, n_2)}(S_1',
\lambda_1')\\
t_2 &= (S_2,
\lambda_2)\pntrans{\{\test{\subeq{x_2}{n_1}},
\map{\subeq{y_2}{n_2}}\}}{\mathit{rec}(n_1, n_2)}(\new r.S_2',
\lambda_2\cdot\lambda_2')\\
t_r&=(\new r.S_2',
\lambda_2\cdot\lambda_2')\pntrans{\{\map{r}{n_{r}}\}}{\tau} (S_2',
\lambda_2\cdot\lambda_2').
\end{align*}
So we assume the communication is on a fresh channel $n_1$ and sends a fresh
name $n_2$. The cases where $x_1$ or $y_1$ are mapped to public names  are
similar.

For marking $M$, the test commands that label transition $(t_1, t_2)$ allow us
to conclude the marking in Line~\eqref{Eq:FromTest}.
\begin{alignat}{5}
M(\subeq{x_1}{n_1})&=1& M(\subeq{x_2}{n_1})&=1&\hspace{0.5cm}
M(\subeq{y_1}{n_2})&=1 \label{Eq:FromTest}\\
M(\subneq{y_2}{n_2})&=1
&\hspace{0.5cm}M(\subeq{y_2}{a})&=0&~\forall
a\in\PUB\cup\NEW.\label{Eq:FromMapping}
\end{alignat}
In the following Line~\eqref{Eq:FromMapping}, the implementation of mapping
requires a token on $\subneq{y_2}{n_2}$.
By \substmarkcompl, this only gives $M(\subeq{y_2}{n_2})=0$.
We derive that actually all $\subeq{y_2}{a}$ are unmarked as follows.
We have $(\class{G}, M)\in\rel$, which means $M$ is known to correspond to a
process.
This process has a substitution that does not contain $y_2$ in its
domain.
This is due to  \domproc\ in combination with the fact that $y_2$ is bound.
Constraint $\cornotdom$ yields $M(\subeq{y_2}{a})=0$ for all
$a\in\PUB\cup\NEW$.
\begin{alignat}{3}
M(\subneq{x}{n_r})&=1~\forall x\in\INP\cup\FRM\cup
\{r_*\}&\hspace{0.5cm}M(\subeq{x}{n_r})&=0~\forall x\in\INP\cup\FRM\cup\RESTR \label{Eq:FromMapRest1}\\
M(\subeq{r}{n})&=0~\forall n\in\NEW.\label{Eq:FromMapRest2}
\end{alignat}
That $t_r$ is enabled gives the first marking in Line~\eqref{Eq:FromMapRest1}.
With \substmarkrest\ and \substmarkcompl, we conclude that no name
maps to $n_r$.
Like for $y_2$, we get that $r$ does not map to any fresh name,
Line~\eqref{Eq:FromMapRest2}.

In the control flow, marking $M'$ differs from $M$ in that
$(S'_1, \lambda_1')$ and $(S_2', \lambda_2\cdot \lambda_2')$
instead of $(S_1, \lambda_1)$ and $(S_2, \lambda_2)$ are marked
in $N(S_{\init, 1})$ and $N(S_{\init, 2})$. For the
substitution net, we only give the places on which the marking
has changed. The following is immediate from the definition of
implementation:
\[
    M'(\subeq{y_2}{n_2})=1
    \qquad
    M'(\subneq{y_2}{n_2})=0
    \qquad
    M'(\subeq{r}{n_r})=1
    \qquad
    M'(\subneq{r_*}{n_r})=0.
\]
$M'=M_1'\cup M_2'$ is stable; moreover, marking $M_1'$ of
$\nsubst$ is indeed a substitution marking.

\smallskip

We now argue that (1.a) there is
$G'\in\factorize{\reachof{F}}{\structCong}$ with
$\class{G}\rightarrowts \class{G}'$ and (1.b) $(\class{G'},
M')\in\rel$.\medskip

\noindent\textbf{Claim 1.a: There is
$G'\in\factorize{\reachof{F}}{\structCong}$ so that
$\class{G}\rightarrowts \class{G}'$}\quad We assume that
$\class{G}$ and $M_1\cup M_2$ are related by $\rel$. Hence,
there is a process in normal form that satisfies
\begin{align*}
G\structCong \new \tilde r.(S_1\sigma_1\parComp S_2\sigma_2\parComp\ldots
\parComp S_{n}\sigma_n).
\end{align*}
From marking $M_1\cup M_2$, we now derive the following information:
\[
    S_1 = \send{x_1}{y_1}.S_1' + \ldots
    \quad
    S_2 = \rec{x_2}{y_2}.\new r.S_2' + \ldots
    \quad
    \sigma_1(x_1)=\sigma_2(x_2)\in \tilde r
    \quad
    \sigma_1(y_1)\in\tilde r.
\]
The equalities on $S_1$ and $S_2$ are due to the markings of the nets
$N(S_{\init, 1})$ and $N(S_{\init, 2})$.
For the substitution, we make use of the fact that $M_1$
corresponds to $\sigma_1\cup\ldots\cup \sigma_n$. We have
$M_1(\subeq{x_1}{n_1})=1=M_1(\subeq{x_2}{n_1})$ with
$n_1\in\NEW$. Since $x_1$ and $x_2$ are bound to at most one
name by \substmarkbind, this allows us to conclude that the
markings of $\subeq{x_1}{n}$ and $\subeq{x_2}{n}$ coincide for
all names $n\in\NEW$. Hence, we get
$\sigma_1(x_1)=\sigma_2(x_2)$ by \coreq. By \domproc, we have
that $\sigma_1(x_1)\in\tilde r\cup \PUB$. If $\sigma_1(x_1)$
was in $\PUB$, we had $M_1(\subeq{x_1}{\sigma_1(x_1)})=1$ by
\corpub. This is not the case, hence $\sigma_1(x_1)\in \tilde
r$. For $y_1$, the reasoning is similar. We already mentioned
above that $\{y_2, r\}\notin \dom(\sigma_1\cup\ldots\cup
\sigma_n)$.

The normal form process has a reaction to
\begin{align*}
G'\DEF \new \tilde r.a_r.(S_1'\sigma_1\parComp
S_2'\sigma_2'\parComp\ldots
\parComp S_n\sigma_n)\quad\text{with}\quad \sigma_2' \DEF
\sigma_2\{\sigma_1(y_1)/y_2\}\{a_r/r\}.
\end{align*}
Hence, $\class{G}\rightarrowts \class{G'}$. Since $G$ was
reachable from $F$, we have $G'$ reachable from $F$. Moreover,
we already argued in the proof of Lemma~\ref{Lemma:ProcToPN}
that $G'$ has the required normal form.\medskip

\noindent\textbf{Claim 1.b: $(\class{G'}, M_1'\cup
M_2')\in\rel$}\quad 
For the threads, the
reasoning is as in
Lemma~\ref{Lemma:ProcToPN}. We check that $M_1'$
corresponds to $\sigma'\DEF\sigma_1\cup\sigma_2'\cup\ldots\cup
\sigma_n$. \cornotdom\ to \correst\ are as before. For \coreq, we consider
$r\neq x\in\dom(\sigma')$. We have $\sigma'(r)\neq \sigma'(x)$, which coincides
with the fact that $M_1'(\subeq{r}{n_r})=1$ and
$M_1'(\subeq{x}{n_r})=0$.
\qed
\medskip

\proof[Proof of Theorem~\ref{Theorem:Bisimulation}] We show
that $\rel$ relates $\class{F}$ and $M_0$. The transitions can
be mimicked due to Lemmas~\ref{Lemma:ProcToPN}
and~\ref{Lemma:PNToProc}.
By our assumptions, we have $F =
\new \tilde r.(S_{\init, 1}\sigma_1\parComp\ldots\parComp
S_{\init, n}\sigma_n)$. Here, $S_{\init, i}$ are choices or
calls that have been derived from artificial defining
equations. Moreover, $\sigma_i:\tilde f_{\init,
i}\rightarrow\tilde r\cup\PUB$ with $ \dom(\sigma_i)=\tilde
f_{\init, i}=\freeNames{S_{\init, i}}\cup\emptyset$. This shows
\domproc, and concludes the proof that $F$ is in normal form.
For the initial marking $M_0=M_{0, 1}\cup M_{0, 2}$ of $N(F)$,
we have that $M_{0, 1}$ corresponds to $\sigma_1\cup\ldots \cup
\sigma_n$ as needed. In the control flow nets $N(S_{\init,
i})$, we have the necessary tokens on $(S_{\init, i},
\varepsilon)$.\qed


\section{Optimisation of the translation}\label{se-opt}

In this section, we propose optimisations of the translation,
which can significantly reduce the size of the resulting safe
PN and increase the efficiency of subsequent model checking.

\subsection{Communication splitting}\label{se-opt-splitting}
Recall the size of the PN resulting from our translation is
dominated by the number of transitions modelling communication.
We now propose a method to significantly decrease this number.
It actually reduces the asymptotic worst case size from
$O(\sizeApp{F}^4)$ down to $O(\sizeApp{F}^3)$. Furthermore, its
straightforward generalisation yields a polynomial translation
from polyadic \picalc to safe PNs, see Sect.~\ref{se-ext}.

The idea is to model the communication between potentially
synchronisable actions $\send{a}{b}$ and $\rec{x}{y}$ not by a
single step but by a pair of steps. The first checks
that $a$ and $x$ are mapped to the same value by the
substitution, and the second maps $y$ to the value of~$b$.

Assume $\send{a}{b}$ and $\rec{x}{y}$ correspond to stub
transitions $t'$ and $t''$. To implement the decomposition, we
create a control place $p_\mathit{middle}$ `in the middle' of
the communication and two sets of transitions, $t^1_i$ and
$t^2_j$. The transitions $t^1_i$, created for each
$i\in\dom(a)\cap\dom(x)$, work as follows. Each $t^1_i$
\begin{iteMize}{$\bullet$}
  \item consumes tokens from the input places of $t'$ and
      $t''$ and produces a token on $p_\mathit{middle}$;
  \item checks by read arcs that $[a=i]$ and $[x=i]$ are
      marked (\ie the substitution maps $a$ and $x$ to the
      same value $i$ and thus the synchronisation is
      possible).
\end{iteMize}
\noindent The transitions $t^2_j$, created for each
$j\in\dom(b)\cap\dom(y)$, work as follows. Each $t^2_j$
\begin{iteMize}{$\bullet$}
  \item consumes a token from $p_\mathit{middle}$ and
      produces tokens on the output places of $t'$
      and~$t''$;
  \item checks by a read arc that $[b=j]$ is marked,
      consumes a token from $[y\neq j]$ (if this place
      exists) and produces a token on $[y=j]$ (mapping thus
      in the substitution $y$ to $j$, \ie to the value of
      $b$).
\end{iteMize}
If the synchronisation is possible in the current state of the
system (\ie $a$ and $x$ have the same value), exactly one of
the transitions $t^1_i$ is enabled; else none of these transitions is
enabled. Once some $t^1_i$ fires, exactly one of the
transitions $t^2_j$ becomes enabled.

\subsection{Abstractions of names}

\newcommand{\abs}       {\mathit{abs}}
\newcommand{\thread}    {\mathit{thread}}
\newcommand{\type}      {\mathit{type}}
\newcommand{\dpth}     {\mathit{depth}}

In the substitution net described in Sect.~\ref{se-subst}, each
bound name and formal parameter is represented by a separate
row of places. In practice, it is often the case that some
bound names and formal parameters can never be simultaneously
active, and so can share the same row of places.

We have implemented a simple sharing scheme by introducing an
equivalence $\sim$ on the set of bound names and formal
parameters, such that if two names are equivalent then they
cannot be simultaneously active. Then the rows of the
substitution table will correspond to the equivalence classes
of $\sim$, and for each bound name and formal parameter we will
introduce the \emph{abstraction} operator $\abs$, mapping the
name to the corresponding equivalence class of $\sim$. Now the
operations on the substitution (initialisation of a restricted
name, remapping and unmapping) can be performed on the
abstractions of names rather than the names themselves.

A possible choice of equivalence $\sim$ and the related
abstraction is as follows. For each name
$b\in\RESTR\cup\INP\cup\FRM$, we denote by $\thread(b)$ the
thread where $b$ is defined (note that due to \noclash{} and
the assumption that threads do not share defining equations,
$\thread(b)$ is unique). Furthermore, we define
\[
    \type(b)
    \DEF
    \left\{
    \begin{array}{rl}
      0 & \mbox{if $b\in\RESTR$} \\
      1 & \mbox{otherwise (\ie if $b\in\INP\cup\FRM$),}
    \end{array}
    \right.
\]
and $\dpth(b)$ to be the number of names
$b'\in\RESTR\cup\INP\cup\FRM$ in whose scope $b$ resides and
such that $\type(b)=\type(b')$. Then the abstraction of $b$ can
be defined as a tuple
\[
    \abs(b)\DEF\Big(\thread(b),\type(b),\dpth(b)\Big),
\]
and two names are considered equivalent \wrt $\sim$ iff their
abstractions coincide. This equivalence ensures that two
distinct names related by it belong to the same thread, and
that their scopes lie within either different defining
equations or different branches of some choice operator, and so
the names cannot be simultaneously active. Other choices of
$\sim$ and $\abs$ are also possible, and we plan to explore
them in our future work.

\subsection{Better overapproximations for name domains}
Recall that the domain of a bound name or formal parameter is
an overapproximation of the set of values from $\PUB\cup\NEW$
that it can take. While the rough overapproximation proposed in
Sect.~\ref{se-translation} is sufficient to make the
translation polynomial, its quality can be improved by static
analysis, resulting in a much smaller PN. In particular, the
number of synchronisations between communication actions as
well as the number of transitions implementing each
communication may be reduced significantly. Furthermore, the
number of transitions implementing parameter passing in calls
and the number of places in $\nsubst$ can also decrease
substantially.

We outline a simple iterative procedure to compute better
overapproximations. We start by setting $\dom(p)\DEF\{p\}$ for
each public name $p$. For each restricted name $r$ we set
$\dom(r)\DEF\{r\}$, interpreting this as that the values of $r$
are taken from the set $\NEW_r$ of unique values (the procedure
never looks inside $\NEW_r$, and only exploits the fact that
the names from $\NEW_r$ are different from all the other
names). The domains of these names are fixed and will not be
changed by the procedure. The domains of all  other names
occurring in the FCP are initialised to $\emptyset$; they will
grow monotonically during the run of the procedure, converging
to some overapproximations.

Each iteration of the procedure identifies two actions,
$\send{a}{b}$ and $\rec{x}{y}$, that satisfy the following
conditions: (i) the actions belong to different threads of the
FCP; (ii) $\dom(a)$ and $\dom(x)$ overlap; and (iii)
$\dom(b)\nsubseteq\dom(y)$. If no two such actions can be
found, the procedure stops, returning the current values of the
domains. Otherwise, $\dom(y)$ is replaced by its union with
$\dom(b)$. Intuitively, the above conditions check that the two
actions can potentially synchronise, and so $y$ can be mapped
to the value of $b$, and so its domain has to include that of
$b$.

\subsection{Better overapproximation for $|\NEW|$}

The cardinality of the set $\NEW$ is an important parameter of the
translation, affecting the efficiency of almost all its
aspects. While the rough overapproximation proposed in
Sect.~\ref{se-translation} (taking $|\NEW|$ to be the total
number of bound names and formal parameters) is sufficient to
make the translation polynomial, a better one can make the
translation much more practical and amenable to model checking.

In the worst case, all names from
$\RESTR\cup\INP\cup\FRM$ can be assigned
different values from $\NEW$. To improve the overapproximation
of Sect.~\ref{se-translation}, we observe that in many cases
not all such names can be simultaneously active, \ie it is
enough to overapproximate the number of such names that can be
simultaneously active. Hence we propose the following improved
overapproximation of $|\NEW|$. If there are no occurrences of
the restriction operator in the FCP, $|\NEW|\DEF 0$. Else, for
each thread we compute the maximal number of names from
$\RESTR\cup\INP\cup\FRM$ that can be simultaneously active in
it,
\begin{align*}
\max\set{\power{\freeNames{S'}\cup \lambda}}{(S, \varepsilon)\text{ is a defining process and }(S, \varepsilon)\pntrans{}{}^* (S', \lambda)},
\end{align*}
and set $|\NEW|$ to the sum of these numbers.

The number of names from $\RESTR\cup\INP\cup\FRM$ that can be
simultaneously active in a thread can be computed by separately
determining this parameter for each of the defining equations
belonging to this thread, as well as the subterm of the main
process corresponding to this thread, and taking the maximum of
these values. Since these \picalc expressions are sequential,
their parse trees can only have the $+$ operator in every node
where a branching occurs, and so the sought value is simply the
maximum of the numbers of active names from
$\RESTR\cup\INP\cup\FRM$ in the leafs of this parse tree.
Furthermore, the names whose domains contain no restricted
names can be ignored by this analysis.

\subsection{Sharing subnets for unmapping names}

When we call $\call{K}{\tilde a}$, some names have to be
unmapped in the substitution. The subnet for unmapping a
particular name can be shared by all points where such
unmapping is necessary. This reduces the size of the resulting
PN. This optimisation is especially effective when name
abstractions (see above) are used, as the sharing increases
significantly in such a case.

\subsection{Re-ordering parameters of calls}

Consider the FCP $\call{K}{a, b}$ with:
\[
\begin{array}{r@{\,}c@{\,}l}
    K(f_1, f_2)&:=&\call{L}{f_2, f_1}\\
    L(f_3, f_4)&:=&\call{K}{f_4,f_3}
\end{array}
\qquad\Rightarrow\qquad
\begin{array}{r@{\,}c@{\,}l}
    K(f_1, f_2)&:=&\call{L}{f_1, f_2}\\
    L(f_4, f_3)&:=&\call{K}{f_4, f_3}
\end{array}
\]
With the definition on the left, when name abstractions are computed,
the equivalence relation has two classes, $\{f_1,f_3\}$ and
$\{f_2,f_4\}$.
Hence, the substitution has to be modified every time the calls
are performed, as the call parameters keep getting flipped. If
the order of the formal parameters in one of the defining
equations is changed (together with the order of the factual
parameters in the corresponding calls), as shown on the right,
the substitution would not require any changes. This significantly reduces the
size of the resulting net.

This example illustrates that the order of formal parameters in
the defining equations matters, and the translation can gain
savings by changing this order. Searching for the best order of
formal parameters can be formulated as an optimisation problem,
with the cost function being the total number of changes
required in the substitution for all the calls.

\subsection{Dropping the restrictions in the main term of the FCP}

All restrictions in the main term of the FCP can
be dropped, making the formerly restricted names public.
Note that due to \noclash{}, this does not introduce name clashes. The transformation
yields a bisimilar \picalc process, but the corresponding PN becomes smaller.

\subsection{Separate pools of values for restricted names}

Creation of names introduces ar\-bit\-ra\-tion between the
values in \NEW (see Sect.~\ref{se-control}): a name that is
currently unused has to be chosen to initialise the given
restricted name. Such arbitration can adversely affect the
efficiency of some model checking methods.

It is possible to eliminate such arbitration by
splitting \NEW into several pools, one for each thread, and initialise restricted names
only from the corresponding pool, by sequentially looking for the
first unused value. This however increases the size of
the resulting PN. Moreover, if symmetry reduction is used in
model checking, the problem vanishes.

\subsection{Using symmetries}

The translation introduces a number of symmetries in the PN:
(i) the values in $\NEW$, and thus the corresponding columns of the substitution (see
Fig.~\ref{fi-subst}), are interchangeable; and
(ii) when enforcing the assumption that threads do not share defining
equations as explained in Sect.~\ref{se-basics}, some equations are replicated.

It is desirable to exploit these symmetries during model checking.
In particular, this would efficiently handle the arbitration that arises when a value from $\NEW$
has to be chosen to initialise a restriction.
If symmetries are used, all the immediate successor states of the arbitration are equivalent, and
only one of them has to be explored further.

\subsection{Translation to different PN classes}

Our translation produces a safe PN, as this PN class is
particularly suitable for algorithmic verification. However, if
the model checking method can cope with more powerful PN
classes, the following changes can be made.\medskip

\noindent\textbf{Translation to bounded PNs}\quad For each
$\mathit{val}\in\NEW$, we can fuse the places
$[\mathit{var}\neq\mathit{val}]$, where
$\mathit{var}\in\{r_*\}\cup\INP\cup\FRM$, into one place
$[*\neq\mathit{val}]$. We thus replace
$|\NEW|\cdot(|\INP|+|\FRM|+1)$ safe places with $|\NEW|$ places
of capacity $|\INP|+|\FRM|+1$. It is still possible to perform
all the necessary operations with the substitution. In
particular, to find a value $\mathit{val}\in\NEW$ to which no
bound name or formal parameter is currently mapped, and map a
given restricted name $r_k$ to $\mathit{val}$, the PN
transition performing the initialisation has to:
\begin{iteMize}{$\bullet$}
  \item consume by a weighted arc $|\INP|+|\FRM|+1$ tokens
      from $[*\neq\mathit{val}]$ (checking thus that
      $\mathit{val}$ is not assigned to any name)
      and return by a weighted arc $|\INP|+|\FRM|$ tokens;
  \item produce a token at $[r_k=\mathit{val}]$.\medskip
\end{iteMize}
\noindent\textbf{Translation to coloured PNs}\quad In this case, the
symmetries present in the PN can be used to fold it. In
particular:
\begin{iteMize}{$\bullet$}
  \item The values in $\NEW$ are interchangeable, and so
      the corresponding columns of the substitution can be
      folded into one column, by giving the tokens
      corresponding to the elements of $\NEW$ unique
      colours.
  \item Instead of enforcing the assumption that the
      threads do not share defining equations (see
      Sect.~\ref{se-basics}), one can use coloured control
      tokens that are unique for each thread.
\end{iteMize}

\section{Extensions}\label{se-ext}

\newcommand{\match}[3] {[#1=#2].#3}
\newcommand{\mismatch}[3] {[#1\neq#2].#3}

We now generalise the translation to some often used extensions
of \picalc{}, \viz to polyadic communication and
synchronisation, and to match and mismatch operators.\medskip

\noindent\textbf{Polyadic communication}\quad Polyadic
communication exchanges multiple names in one reaction.
Intuitively, a sending prefix $\send{a}{x_1\ldots x_n}$ and a
receiving prefix $\rec{b}{y_1\ldots y_n}$ (with all $y_i$ being
different names) can synchronise iff $\sigma(a)=\sigma(b)$.
After synchronisation each $y_i$ gets the value of $x_i$.
Formally,
\begin{align*}
    \rec{a}{\tilde y}.S_1+\ldots\parComp \send{a}{\tilde x}.S_2+\ldots\rightarrow S_1\{\tilde x/ \tilde y\} \parComp S_2\quad\text{if }|\tilde x|=|\tilde y|.
\end{align*}

A polynomial translation of this extension generalises the
communication splitting idea described in Sect.~\ref{se-opt}.
We perform the communication in stages.
At the first step, one checks that $a$ and $b$ are mapped to the
same value by the substitution. The subsequent
steps map, one-by-one, $y_i$ to the value of $x_i$ in the
substitution.
\medskip

\noindent\textbf{Polyadic synchronisation}\quad Dual to
polyadic communication is polyadic
synchronisation~\cite{CarboneMaffeis2003}. When sending a
message, this operation synchronises on multiple channels
instead of just one. Formally,
\[
    \rec{\tilde a}{y}.S_1+\ldots\parComp\send{\tilde a}{x}.S_2+\ldots\rightarrow S_1\{x/y\}\parComp S_2.
\]
Polyadic synchronisation captures, in a clean formalism,
expressive features like locality. To extend our translation to
polyadic synchronisation, we check, one-by-one, that the
channel bindings of both processes match.

In the presence of polyadic synchronisation, the relationship
between the FCP and its PN translation is subtle, since false
deadlocks may be introduced. For example, in $\rec{a_1\cdot
a_2}{y}\pc \send{b_1\cdot b_2}{x}$ the evaluation may find
$a_1$ and $b_1$ bound to the same name,
$\sigma(a_1)=\sigma(b_1)$, while $a_2$ and $b_2$ do not match.
In this case the resulting PN will get stuck in the middle of
the evaluation. This does not happen in the original \picalc
process. Nevertheless, such false deadlocks can easily be
distinguished from real ones, and so the resulting PN is still
suitable for model checking. An alternative is to use the idea
of the construction in Sect.~\ref{Section:PNtoFCPtranslation},
which avoids false deadlocks.\medskip

\noindent\textbf{Match and mismatch operators}\quad The match
and mismatch operators are a common extension of \picalc.
Intuitively, the process $\match{x}{y}{S}$ behaves as $S$ if
$\sigma(x)=\sigma(y)$ and does nothing otherwise, and the
process $\mismatch{x}{y}{S}$ behaves as $S$ if $\sigma(x)\neq
\sigma(y)$ and does nothing otherwise. To handle these
operators, we extend the construction of $N(S_{\init})$ with
the following transitions. For each $a\in\PUB\cup\NEW$, we have
\[
    (\match{x}{y}{S},\lambda)
    \pntrans{\{\test{\subeq{x}{a}},\test{\subeq{y}{a}}\}}{\tau}
    (S,\lambda\cdot \lambda')
    \qquad
    (\mismatch{x}{y}{S},\lambda)
    \pntrans{\{\test{\subeq{x}{a}},\test{\subneq{y}{a}}\}}{\tau}
    (S,\lambda\cdot \lambda'),
\]
where $\lambda'$ contains the names from
$\{x,y\}\setminus\freeNames{S}$. For the latter rule, new
places $\subneq{x}{a}$ complementing $\subeq{x}{a}$ may have to
be introduced to $\nsubst$. The relationship between FCP and PN
is similar to the case of polyadic synchronisation.


%
%
%

\section{Experimental results}\label{se-experiments}

\newcommand{\CS}{\textit{CS}}
\newcommand{\CLIENT}{\mathit{CLIENT}}
\newcommand{\SERVER}{\mathit{SERVER}}
\newcommand{\myURL}{\mathit{url}}
\newcommand{\ip}{\mathit{ip}}
\newcommand{\session}{\mathit{s}}
\newcommand{\ses}{\mathit{ses}}
\newcommand{\SESSION}{\mathit{SESSION}}
\newcommand{\getSES}{\mathit{getses}}

\newcommand{\NESS}{\textit{NESS}\xspace}
\newcommand{\DETNESS}{\textit{DNESS}\xspace}
\newcommand{\TEACHER}{\mathit{T}}
\newcommand{\STUDENT}{\mathit{S}}
\newcommand{\studChan}{\mathit{h}}
\newcommand{\nessChan}{\mathit{nessc}}
\newcommand{\nsc}{\mathit{nsc}}
\newcommand{\done}{\mathit{fin}}
\newcommand{\ENV}{\mathit{ENV}}

\newcommand{\GSM}{\textit{GSM}\xspace}
\newcommand{\PHONES}{\textit{PHONES}\xspace}

\newcommand{\fcptopn}   {\textsc{Fcp2Pn}\xspace}
\newcommand{\lola}   {\textsc{LoLA}\xspace}

To demonstrate the practicality of our translation-based approach to
\picalc\ verification, we implemented
the encoding of FCPs into safe PNs in the tool
\fcptopn\footnote{
    Available from
    \url{http://homepages.cs.ncl.ac.uk/victor.khomenko/tools/fcp2pn}.
    },
and used the resulting nets for model checking a number of
benchmark systems.\footnotemark[\value{footnote}]

The \NESS \emph{(Newcastle E-Learning Support System)} series
of benchmarks models an electronic coursework submission
system \cite{KKN2006}. The model consists of a
teacher process $\TEACHER$ composed in parallel with $k$
students $\STUDENT$ (the system can be scaled up by increasing
the number of students) and an environment process $\ENV$.
Every student has its own local channel for communication,
${\studChan}_i$, and all students share the channel
$\studChan$:
\[
    \new {\studChan}.\new {\studChan}_1\ldots\new {\studChan}_k.
    \Big(
        \call{\TEACHER}{\nessChan, {\studChan}_1,\ldots,{\studChan}_k}
        \pc
        \prod_{i=1}^k\call{\STUDENT}{{\studChan}, {\studChan}_i}
        \pc
        \call{\ENV}{\nessChan}
    \Big)\;.
\]
The students are supposed to submit their work for assessment
to \NESS. The teacher passes the channel $\nessChan$ of the
system to all students, $\send{h_i}{\nessChan}$, and then waits
for the confirmation that they have finished working on the
assignment, $\rec{h_i}{x_i}$. After receiving the \NESS
channel, $\rec{{\studChan}_i}{\nsc}$, students
non-deterministically organise themselves in pairs. To do so,
they send their local channel ${\studChan}_i$ on $\studChan$
and at the same time listen on $\studChan$ to receive a
partner,
$\send{\studChan}{{\studChan}_i}\ldots+\rec{\studChan}{x}\ldots$
When they finish, exactly one student of each pair sends two
channels (the own channel ${\studChan}_i$ and the channel
received from the partner) to the support system,
$\send{\nsc}{{\studChan}_i}.\send{\nsc}{x}$, which give access
to their completed joint work. These channels are received by
the $\ENV$ process. The students finally notify the teacher
about the completion of their work,
$\send{{\studChan}_i}{\done}$. Thus, the system is modelled by:
\begin{align*}
\TEACHER(\nessChan, {\studChan}_1,\ldots, {\studChan}_k):=&\prod^k_{i=1}\send{{\studChan}_i}{\nessChan}.\rec{{\studChan}_i}{x_i}.\stopProc\\
\STUDENT({\studChan}, {\studChan}_i):=&\rec{{\studChan}_i}{\nsc}.(\send{\studChan}{{\studChan}_i}.\send{{\studChan}_i}{\done}.\stopProc+
                                      \rec{\studChan}{x}.\send{\nsc}{{\studChan}_i}.\send{\nsc}{x}.\send{{\studChan}_i}{\done}.\stopProc)\\
\ENV(\nessChan):=&\rec{\nessChan}{y_1}.~\ldots~.\rec{\nessChan}{y_k}.\stopProc
\end{align*}
To distinguish proper termination from deadlocks (where
some processes are stuck waiting for a communication),
a new transition is added to the PN that creates a loop at the state corresponding to
successful termination. Obviously, the system
successfully terminates iff the number of students is even, \ie
they can be organised into pairs.

The \DETNESS model is a refined version of \NESS, with
deterministic pairing of students. Thus, the number of students
is always even, and these benchmarks are deadlock-free.

The \CS($m$,$n$) series of benchmarks models a client-server
system with one server, $n$ clients, and the server spawning
$m$ sessions that handle the clients' requests:
\begin{align*}
\CLIENT(\myURL)&:=\new\ip.\send{\myURL}{\ip}.\rec{\ip}{\session}.\rec{\session}{x}.\call{\CLIENT}{\myURL}\\
\SERVER(\myURL,\getSES)&:=\rec{\myURL}{y}.\rec{\getSES}{s}.\send{y}{s}.\call{\SERVER}{\myURL,\getSES}\\
\SESSION(\getSES)&:=\new \ses.\send{\getSES}{\ses}.\send{\ses}{\ses}.\call{\SESSION}{\getSES}\\[-2.2em]
\end{align*}
\[
\new\getSES\Big(\SERVER(\myURL,\getSES)\pc\prod_{i=1}^m \SESSION(\getSES)\pc\prod_{i=1}^n \CLIENT(\myURL)\Big)
\]
On a client's request, the server creates a new session using
the $\getSES$ channel, $\rec{\getSES}{s}$. A session is
modelled by a $\SESSION$ process. It sends its private channel
$\new\ses$ along the $\getSES$ channel to the server. The
server forwards the session to the client, $\send{y}{s}$, which
establishes the private session, and becomes available for
further requests. A communication on  the channel $\ses$ terminates the private session.
All these benchmarks are deadlock-free.

The \GSM benchmark is a specification of the handover procedure
in the GSM Public Land Mobile Network. We use the well-known
\picalc model from~\cite{OravaParrow1992}, with one mobile
station, two base stations, and one mobile switching. We also
studied a variant \GSM' where a restriction in the sender
process is dropped: the sender keeps sending the same message
instead of generating a new one every time. Since the content
of the message is not important, this change is inconsequential
from the modelling point of view. However, it significantly
reduces the size of the PN. Indeed, the modified FCP is
restriction-free, and so $\NEW=\emptyset$.

The \PHONES benchmark is a classical example taken
from~\cite{Milner1999}, modelling a handover procedure for
mobile phones communicating with fixed transmitters, where the
phones have to switch their transmitters on the go.

\begin{table}[t]
    \centering
    \scriptsize
\begin{tabular}{|l|rrr|rr|r|}
  \hline
  \multicolumn{1}{|c|}{} & \multicolumn{3}{c|}{Process size} & \multicolumn{2}{c|}{Safe PN} & \multicolumn{1}{c|}{Dlck}\\
  \multicolumn{1}{|c|}{Problem} & \multicolumn{1}{c}{FCP} & \multicolumn{1}{c}{nfFCP} & \multicolumn{1}{c|}{$|\NEW|$} & \multicolumn{1}{c}{$|P|$} & \multicolumn{1}{c|}{$|T|$} & \multicolumn{1}{c|}{[sec]}\\
  \hline
  \NESS(04) & 110 & 110 & 0 & 137 & 145 & 0.02\\
  \NESS(05)\dag & 137 & 137 & 0 & 196 & 246 & 0.09\\
  \NESS(06) & 164 & 164 & 0 & 265 & 385 & 0.16\\
  \NESS(07)\dag & 191 & 191 & 0 & 344 & 568 & 0.45\\
  \hline
  \DETNESS(06) & 118 & 118 & 0 & 157 & 103 & 0.02\\
  \DETNESS(08) & 157 & 157 & 0 & 241 & 169 & 0.05\\
  \DETNESS(10) & 196 & 196 & 0 & 341 & 251 & 0.13\\
  \DETNESS(12) & 235 & 235 & 0 & 457 & 349 & 2.27\\
  \DETNESS(14) & 274 & 274 & 0 & 589 & 463 & 1.71\\
  \hline
\end{tabular}%
\quad%
\begin{tabular}{|l|rrr|rr|r|}
  \hline
  \multicolumn{1}{|c|}{} & \multicolumn{3}{c|}{Process size} & \multicolumn{2}{c}{Safe PN} & \multicolumn{1}{c|}{Dlck}\\
  \multicolumn{1}{|c|}{Problem} & \multicolumn{1}{c}{FCP} & \multicolumn{1}{c}{nfFCP} & \multicolumn{1}{c|}{$|\NEW|$} & \multicolumn{1}{c}{$|P|$} & \multicolumn{1}{c|}{$|T|$} & \multicolumn{1}{c|}{[sec]}\\
  \hline
  \CS(2,1) & 45 & 54 & 7 & 138 & 149 & 1.01\\
  \CS(2,2) & 48 & 68 & 10 & 243 & 320 & 0.16\\
  \CS(3,2) & 51 & 80 & 11 & 284 & 431 & 1.28\\
  \CS(3,3) & 54 & 94 & 14 & 428 & 728 & 3.67\\
  \CS(4,4) & 60 & 120 & 18 & 663 & 1368 & 11.73\\
  \CS(5,5) & 66 & 146 & 22 & 948 & 2288 & 46.61\\
  \hline
  \GSM & 175 & 231 & 12 & 636 & 901 & 4.39\\
  \GSM' & 174 & 230 & 0 & 355 & 503 & 3.09\\
  \hline
  \PHONES & 157 & 157 & 0 & 131 & 94 & 0.01\\
  \hline
\end{tabular}
\vspace*{0.5em}
    \caption{\label{ta-exp}Experimental results.}
\end{table}

The experimental results are given in Table~\ref{ta-exp}, with
the columns showing from left to right: name of the case
study (\dag{} indicates deadlocks); sizes of the original FCP
and its normal form (see Sect.~\ref{se-basics}), together with the
cardinality of \NEW determined by static analysis; number of
places and transitions in the resulting safe PN; and deadlock
checking time.

The experiments were conducted on a PC with an Intel Core 2
Quad Q9400 2.66 GHz processor (a single core was
used) and 4G RAM\@. The deadlock checking was performed with
the \lola tool,\footnote{
    Available from
    \url{http://service-technology.org/tools/lola}.
    }
configured to assume safeness of the PN (\texttt{CAPACITY
1}), use the stubborn sets and symmetry reductions
(\texttt{STUBBORN}, \texttt{SYMMETRY}), compress states using
P-invariants (\texttt{PREDUCTION}), use a light-weight data
structure for states (\texttt{SMALLSTATE}), and check for
deadlocks (\texttt{DEADLOCK}). The FCP to PN translation times
were negligible ($<0.1$ sec in all cases) and so are not
reported.

The experiments indicate that the sizes of the PNs grow moderately with the
sizes of the FCPs, and that the PNs are suitable for efficient verification:
deadlock checking took less than a minute in all examples.

\section{PN to FCP translation}\label{Section:PNtoFCPtranslation}

\newcommand{\sd}[1] {\overline{#1}}
\newcommand{\rc}[1] {#1}

\newcommand{\fail}{{\scriptscriptstyle\times}}
\newcommand{\reset}{\mathit{reset}}
\newcommand{\go}{\mathit{go}}
\newcommand{\success}{\mathit{success}}
\newcommand{\failure}{\mathit{failure}}
\newcommand{\PLACE}{\mbox{\textsc{Empty}}}
\newcommand{\MPLACE}{\mbox{\textsc{Marked}}}
\newcommand{\TRAN}{\mbox{\textsc{Tran}}}
\newcommand{\TRYCONS}{\mbox{\textsc{TryCons}}}
\newcommand{\RETURNTOKENS}{\mbox{\textsc{Ret}}}
\newcommand{\PROD}{\mbox{\textsc{Prod}}}
\newcommand{\SCHEDULER}{\mbox{\textsc{Scheduler}}}

We now study a translation in the reverse direction. We
translate safe PNs into FCPs that are weakly bisimilar and thus
do not introduce false deadlocks. Moreover, no livelocks are
introduced. This improves over the translations
in~\cite{AM2002,Meyer2009Thesis}.

The motivation for proposing this translation is twofold.
First, we obtain a \pspace{} lower bound for verification
problems on FCPs, see Sect.~\ref{se-complexity_intro}. Second,
the idea can be adapted to translate polyadic communication and
match/mismatch operators into safe PNs in a faithful way, see
Sect.~\ref{se-ext}.

The main difficulty is faithful modelling of $n$-ary
synchronisations in PNs by a sequence of binary
synchronisations in \picalc. We address it in three steps.
First, we give a folklore translation of PNs into
FCPs~\cite{AM2002,Meyer2009Thesis}, whose advantage is its
simplicity. The drawback is that it can introduce false
deadlocks. To fix this problem, we develop a second
translation, which yields a weakly bisimilar FCP. This in
particular implies the absence of false deadlocks. The
translation may, however, introduce livelocks. We eliminate
these livelocks in our third translation using scheduling.

All these translations are linear. Moreover, they use a very
restricted set of \picalc capabilities: communications do not
pass information (in particular, no reference passing is used),
no restricted names are used, the calls do not have any
parameters, no $\tau$ actions are used, and the result is a
safe FCP (see Sect.~\ref{se-basics}). As a consequence, the
translations can be adopted to process calculi with weaker
communication capabilities, such as CCS~\cite{Milner1980}.

We fix $N=(P,T,F,M_0)$ as the safe PN to be translated. All
communications will pass a fixed public name $\varepsilon$,
which is not used for any other purpose. Thus, the simplified
syntax $\sd{a}$ and $\rc{x}$ is used instead of
$\send{a}{\varepsilon}$ and $\rec{x}{y}$, respectively.
Similarly, since calls do not pass parameters, we write $K$ for
$\call{K}{-}$.\medskip

\noindent\textbf{Blocking translation}\quad The following
translation is inspired by~\cite{AM2002,Meyer2009Thesis}. For
each place $p\in P$, there is a separate public channel, also
denoted by $p$, and a thread with two defining equations
corresponding to the presence and absence of a token in $p$:
\begin{align*}
    \MPLACE_p:=&\ \sd{p}.\PLACE_p\\
     \PLACE_p:=&\ \rc{p}.\MPLACE_p
\end{align*}
A marked place can send a message over channel $p$,
which models token consumption, and become empty.
Similarly, an empty place can receive a message over $p$, which models
token production, and become marked.

For each transition $t\in T$ with
$\pre{t}=\{p_1,p_2,\ldots,p_m\}$ and
$\post{t}=\{q_1,q_2,\ldots,q_n\}$, we
create a thread with the following defining equation:
\begin{align*}
    \TRAN_t:=\rc{p_1}.\rc{p_2}.\;\ldots\;.\rc{p_m}.\sd{q_1}.\sd{q_2}.\;\ldots\;.\sd{q_n}.\TRAN_t
\end{align*}
Intuitively, the transition consumes tokens, one-by-one, from
the places in $\pre{t}$ by receiving messages over the
corresponding channels. Then it produces tokens, one-by-one, in
the places in $\post{t}$, by sending messages over the
corresponding channels. Since the PN is safe, it is guaranteed
that the places in $\post{t}$ were empty.

The initial term of the FCP is as follows:
\begin{align*}
    \prod_{p\in M_0}\!\! \MPLACE_p
    \parComp
    \!\!\!\!\!\prod_{p\in P\setminus M_0}\!\!\!\!\! \PLACE_p
    \parComp
    \prod_{t\in T} \TRAN_t
\end{align*}
Clearly, the size of the FCP is linear in the size of the
original PN. However, this basic translation can introduce
false deadlocks. Indeed, it is possible for a thread $\TRAN_t$
to consume some, but not all tokens from $\pre{t}$, and become
blocked. In such a case the already consumed tokens are not
returned back to the corresponding places in $\pre{t}$, which
can prevent other transitions from firing. In practice, one can
deal with this problem by declaring the process in which some
$\TRAN_t$ is not in the beginning of its control flow as
unstable, and considering only stable processes.\medskip

\noindent\textbf{Non-blocking translation}\quad To fix the
blocking translation, we change the specification of the place
and transition processes. The idea is to let transitions detect
that some tokens in the preset are missing, and return the
already consumed tokens in such a case.

The initial term is the same as above.
For each place $p\in P$ we create another public channel $p^\fail$, over
which the place process can communicate a
failure to consume a token if it is empty. The
specification of the place process is amended as follows:
\begin{align*}
     \PLACE_p:=\rc{p}.\MPLACE_p + \sd{p^\fail}.\PLACE_p
\end{align*}
The definition of $\MPLACE_p$ remains the same.

The specification of the transition process is now as follows:
\begin{align*}
    \TRAN_t:=&\ \TRYCONS_t^1\\[0.5em]
    \TRYCONS_t^1:=&\ \rc{p_1}.\TRYCONS_t^2+\rc{p_1^\fail}.\RETURNTOKENS_t^0\\
    \TRYCONS_t^2:=&\ \rc{p_2}.\TRYCONS_t^3+\rc{p_2^\fail}.\RETURNTOKENS_t^1\\
    \ldots\\
    \TRYCONS_t^m:=&\ \rc{p_m}.\PROD_t + \rc{p_m^\fail}.\RETURNTOKENS_t^{m-1}\\[0.5em]
    \RETURNTOKENS_t^0:=&\ \TRAN_t\\
    \RETURNTOKENS_t^1:=&\ \sd{p_1}.\RETURNTOKENS_t^0\\
    \ldots\\
    \RETURNTOKENS_t^{m-1}:=&\ \sd{p_{m-1}}.\RETURNTOKENS_t^{m-2}\\[0.5em]
    \PROD_t:=&\ \sd{q_1}.\sd{q_2}.\;\ldots\;.\sd{q_n}.\TRAN_t
\end{align*}
Initially, $\TRAN_t$ tries to consume all the tokens from
$\pre{t}=\{p_1,p_2,\ldots,p_m\}$, one-by-one, by calling
$\TRYCONS_t^1$. $\TRYCONS_t^i$ communicates with the thread for
$p_i$ and either consumes a token from this place (by receiving
a message from channel $p_i$) and then calls
$\TRYCONS_t^{i+1}$ to consume the remaining tokens, or detects
that the place is empty (by receiving a message over channel
$p_i^\fail$), in which case it calls $\RETURNTOKENS_t^{i-1}$ to
return the previously consumed tokens, if any.
Note that the tokens are returned in the reverse order of their consumption.
Once all tokens have been successfully consumed,
$\PROD_t$ is called to communicate with the processes for the
places in $\post{t}=\{q_1,q_2,\ldots,q_n\}$, one-by-one, to
produce tokens on them.

The size of the FCP is still linear in the size of the original
PN, and there are no false deadlocks. In fact, this
FCP is weakly bisimilar to the original PN. It may, however,
introduce livelocks, \eg a disabled transition can perpetually
try and fail to fire.\medskip

\noindent\textbf{Scheduling translation}\quad To solve the
problem of livelocks, we augment the non-block\-ing translation
with the process $\SCHEDULER$ responsible for the global
operation of the net. Hence, the initial term of the FCP
becomes
\begin{align*}
    \prod_{p\in M_0}\!\! \MPLACE_p
    \parComp
    \!\!\!\!\!\prod_{p\in P\setminus M_0}\!\!\!\!\! \PLACE_p
    \parComp
    \prod_{t\in T} \TRAN_t
    \parComp
    \SCHEDULER
\end{align*}
The simulation is performed in rounds. In each round the
scheduler tries to execute a single transition, trying them
one-by-one in some non-deterministically chosen order. If none
of them can be executed, the scheduler blocks. Otherwise, some
transition is executed and a new round starts.

Assuming that $T=\{t_1,t_2,\ldots,t_k\}$, the specification of
the scheduler is as follows:
\begin{align*}
    \SCHEDULER:=\sd{\go}.(\rc{\failure}.\SCHEDULER + \rc{\success}.\sd{\reset_{t_1}}.\sd{\reset_{t_2}}.\;\ldots\;.\sd{\reset_{t_k}}.\SCHEDULER)
\end{align*}
In each round, the scheduler non-deterministically
chooses a transition by communicating with a process $\TRAN_t$
over the public channel $\go$.
Upon receiving a message from the selected transition about a failed
execution attempt, the scheduler chooses another available
transition, or blocks if there is none (in which case the
simulated PN has reached a deadlock state). Upon receiving a
message about successful execution of a transition, the
scheduler communicates over channels $\reset_t$ with every
transition process in some fixed order to make them available
again. The transition processes that are still available also
participate in this communication, but do nothing in response.
Then a new round starts.

The specification of the place process remains the same, and
that of the transition process is amended as follows:
\begin{align*}
    \TRAN_t:=&\ \rc{\reset_t}.\TRAN_t + \rc{go}.\TRYCONS_t^1\\
    \RETURNTOKENS_t^0:=&\ \sd{\failure}.\rc{\reset_t}.\TRAN_t\\
    \PROD_t:=&\ \sd{q_1}.\sd{q_2}.\;\ldots\;.\sd{q_n}.\sd{\success}.\TRAN_t
\end{align*}
The definitions of $\TRYCONS_t^i$ and $\RETURNTOKENS_t^j$ for
$j\neq 0$ remain the same.

Initially, $\TRAN_t$ may receive a reset request over the
public channel $\reset_t$ and return to the initial
state. Alternatively, it competes with the other transitions to
communicate over the $\go$ channel, which results in a
non-deterministic selection of a transition to be executed.
Note that the non-deterministic selection is essential here,
as any fixed order of selection may perpetually ignore some
enabled transition by always executing preceding
transitions. The selected transition process then works as in
the non-blocking translation above, but additionally
communicates its failure or success to the scheduler.
In case of failure, the transition is blocked until it is reset by the scheduler
with a communication over $\reset_t$.

One can see that the resulting FCP is weakly bisimilar with the
original PN, and the problem of livelocks is solved.

\section{Conclusions}\label{se-concl}

We developed a polynomial translation from finite control
\picalc\ processes to safe low-level Petri nets. To our
knowledge, this is the first such translation. There is a close
correspondence between the control flow of the \picalc
specification and the resulting PN, and the latter is suitable
for practical model checking. The translation has been
implemented in the \fcptopn tool, and the experimental results
are encouraging.

We have also proposed a number of optimisations allowing one to
reduce the size of the resulting PN. Moreover, we have shown
how to generalise the translation to more expressive classes of
processes. In particular, we discussed how to handle polyadic
communication, polyadic synchronisation, and match/mismatch
operators.

In future work, we plan to further improve the translation by a
more thorough static analysis, and to incorporate it into
different model checking tool-chains, in particular, ones based
on PN unfolding prefixes and abstraction-refinement approaches.
Moreover, it would be interesting to check if some analog of
Theorem~\ref{Theorem:Bisimulation} holds for barbed
bisimulation. Furthermore, we would like to consider the
labelled semantics of \picalc.

\medskip

\noindent\textbf{Acknowledgements}\quad We would like to thank
Ivan Poliakov for his help with the experiments, and the
anonymous reviewers for their helpful comments.

\bibliographystyle{plain}
\bibliography{literatur}
\end{document}

%% file: pics/substitution_net.tex
\begin{figure}[!t]
    \centering
    \small
    \begin{tabular}{c|llc|llc}
              & \multicolumn{1}{c}{$p_1$} & \multicolumn{1}{c}{$p_2$} & \ldots & \multicolumn{1}{c}{$n_1$} & \multicolumn{1}{c}{$n_2$} & \ldots\\
        \hline
        $i_1$ & \subeqnm{i_1}{p_1} & \subeqnm{i_1}{p_2} & \ldots & \subeqnm{i_1}{n_1} & \subeqm{i_1}{n_2} & \ldots\\
              &  &  &  &  \subneqm{i_1}{n_1} & \subneqnm{i_1}{n_2} & \ldots\\
         $i_2$ & \subeqnm{i_2}{p_1} & \subeqnm{i_2}{p_2} & \ldots & \subeqnm{i_2}{n_1} & \subeqnm{i_2}{n_2} & \ldots\\
              &  &  &  & \subneqm{i_2}{n_1} & \subneqm{i_2}{n_2} & \ldots\\
        \vdots & \multicolumn{1}{c}{\vdots} & \multicolumn{1}{c}{\vdots} & \multicolumn{1}{c|}{$\ddots$} & \multicolumn{1}{c}{\vdots} & \multicolumn{1}{c}{\vdots} & $\ddots$\\
        \hline
        $f_1$ & \subeqnm{f_1}{p_1} & \subeqnm{f_1}{p_2} & \ldots & \subeqm{f_1}{n_1} & \subeqnm{f_1}{n_2} & \ldots\\
              &  &  &  &  \subneqnm{f_1}{n_1} & \subneqm{f_1}{n_2} & \ldots\\
         $f_2$ & \subeqnm{f_2}{p_1} & \subeqnm{f_2}{p_2} & \ldots & \subeqnm{f_2}{n_1} & \subeqnm{f_2}{n_2} & \ldots\\
              &  &  &  &  \subneqm{f_2}{n_1} & \subneqm{f_2}{n_2} & \ldots\\
        \vdots & \multicolumn{1}{c}{\vdots} & \multicolumn{1}{c}{\vdots} & \multicolumn{1}{c|}{$\ddots$} & \multicolumn{1}{c}{\vdots} & \multicolumn{1}{c}{\vdots} & $\ddots$\\
        \hline
        $r_1$ & \multicolumn{3}{c|}{restricted names are never mapped} & \subeqnm{r_1}{n_1} & \subeqnm{r_1}{n_2} & \ldots\\
        $r_2$ & \multicolumn{3}{c|}{to public ones, so no places here} & \subeqnm{r_2}{n_1} & \subeqnm{r_2}{n_2} & \ldots\\
        \vdots & \multicolumn{3}{c|}{} & \multicolumn{1}{c}{\vdots} & \multicolumn{1}{c}{\vdots} & \multicolumn{1}{c}{$\ddots$}\\
              & \multicolumn{3}{c|}{} & \multicolumn{3}{c}{}\\
              &  &  &  & \subneqm{r_*}{n_1} & \subneqm{r_*}{n_2} & \ldots\\
    \end{tabular}
    \caption{\label{fi-subst}
        Illustration of $\nsubst$ with a substitution
marking that corresponds to $\sigma:\{i_1, f_1\}\rightarrow \tilde r\cup\PUB$
where $\sigma(i_1)=r_1$ and $\sigma(f_1)=r_2$ with $r_1\neq r_2$. The
marking represents $r_1$ by $n_2$ and $r_2$ by
$n_1$.}
\end{figure}